\documentclass[rapids]{jfm}
\usepackage{amsmath,amssymb,bm}
\usepackage[usenames,dvipsnames]{color}
\usepackage{natbib}
\usepackage{graphicx}
\usepackage{gensymb}
\usepackage{bm}
\usepackage{amssymb}	
\usepackage[normalem]{ulem}

\newcommand\Rec{\Rey_c}
\usepackage{color}
\newcommand{\MC}[1]{{\color{black}#1}}

\newcommand{\DB}[1]{{\color{black}#1}}

\begin{document}

\title{Universal continuous transition to turbulence in a planar shear flow}
\shorttitle{Continuous transition in a planar shear flow}

\author[M. Chantry, L.~S. Tuckerman and D. Barkley]{
Matthew Chantry$^{1,2,4}$, 
Laurette S. Tuckerman$^{2,4}$ 
and \\ Dwight Barkley$^{3,4}$
}

\affiliation{
$^1$Atmospheric, Oceanic and Planetary Physics, University of Oxford, 
Clarendon Laboratory, Parks Road, Oxford OX1 3PU, UK\\
[\affilskip] $^2$Laboratoire de Physique et M\'ecanique des Milieux 
H\'et\'erog\`enes (PMMH),
CNRS, ESPCI Paris, PSL Research University; Sorbonne Universit\'e, 
Univ. Paris Diderot, France \\
[\affilskip] $^3$Mathematics Institute, University of Warwick, 
Coventry CV4 7AL, UK \\
[\affilskip] $^4$Kavli Institute for Theoretical Physics, 
University of California at Santa Barbara, Santa Barbara, CA 93106, USA
}

\maketitle

\date{\today}

\begin{abstract}

We examine the onset of turbulence in Waleffe flow -- the planar shear flow
between stress-free boundaries driven by a sinusoidal body force.  By
truncating the wall-normal representation to four modes, we are able to
simulate system sizes an order of magnitude larger than
any previously simulated, and thereby to attack the question of universality
for a planar shear flow.  We demonstrate that the equilibrium turbulence
fraction increases continuously from zero above a critical Reynolds number and
that statistics of the turbulent structures exhibit the power-law scalings of
the \MC{(2+1)D} directed percolation universality class.

\end{abstract}

\section{Introduction}

The transition to turbulence in wall-bounded shear flows has been studied for
well over a century, and yet, only recently have experiments, numerical
simulations, and theory advanced to the point of providing a
comprehensive understanding of the route to turbulence in such flows.
Of late, research has focused on how turbulence first appears and becomes
sustained.
The issue is that typically wall-bounded shear flows undergo subcritical
transition, meaning that as the Reynolds number is increased, turbulence does
not arise through \MC{a linear} instability of laminar flow, but instead
appears directly as a highly nonlinear state.
Moreover, the flow does not simply become everywhere turbulent beyond a
certain Reynolds number.  Rather, turbulence initially appears as 
localised patches interspersed within laminar flow. The resulting flow takes
on a complex spatiotemporal form with competing turbulent and laminar
domains. This, in turn, greatly complicates the quantitative analysis of
turbulent transition in subcritial shear flows.
See \cite{Barkley_Theoretical_2016} and \cite{Manneville_Transition_2016} for
recent reviews.

In the 1980s the connection was developed between spatially extended
dynamical systems and subcritical turbulent flows. This provided a broad and
useful context in which to view turbulent-laminar intermittency.
\cite{Kaneko_Spatiotemporal_1985}  constructed minimal models that 
demonstrated how dynamical systems with chaotic
(``turbulent'') and steady (``laminar'') phases would naturally generate
complex spatiotemporal patterns.  Simple models were further studied
by \cite{Chate_Spatio_1988} amongst others.
%
%
\DB{ At the same time, \cite{pomeau1986front} observed that subcritical fluid
  flows have the characteristics of non-equilibrium systems exhibiting what is
  known as an absorbing state transition. Based on this, he postulated that
  these flows might fall into the universality class of directed
  percolation. This would imply that the turbulence fraction varies
  continuously with Reynolds number, going from zero to non-zero at a
  critical Reynolds number, with certain very specific power laws holding at
  the onset of turbulence. (These concepts will be explained further in
  \S\ref{sec:CML}.)}
%
%
%
Since then considerable effort has been devoted to investigating these issues.
The first experimental observation of directed percolation was reported by
\cite{takeuchi2007directed,takeuchi2009experimental} for electroconvection in
nematic liquid crystals.

The status of our understanding for prototypical subcritical shear flows is as
follows.
For pipe flow, there are extensive measurements of the localized turbulent
patches (puffs) that drive the transition to turbulence and we have a good
estimate of the critical point for the onset of sustained turbulence
\citep{avila2011onset}.  However, currently there is no experimental or
computational measurement of the scalings from which to determine whether the
flow is, or is not, in the universality class of directed percolation,
although model systems support that the transition is in this class
\citep{barkley11,Shih_Ecological_2016,Barkley_Theoretical_2016}.
The scaling exponents depend on the spatial dimension of the system. 
\cite{lemoult2016directed} recently carried out a study of 
Couette flow highly confined in two directions so that large-scale
turbulent-laminar intermittency could manifest itself only along one spatial
dimension.  
%
%
\DB{In both experiments and numerical simulations, they measured turbulence
  fraction as a function of Reynolds number and analysed the spatial and
  temporal correlations close to the critical Reynolds number. The results
  support a continuous variation of the turbulence fraction, from zero to
  non-zero at the onset of turbulence, with scaling laws consistent with the
  expectations for directed percolation in one spatial dimension.  }

In systems in which the flow is free to evolve in two large spatial
directions, such as 
Couette and channel flow, the problem is much more
difficult and the situation is less clear.
%
%
Past work has suggested that the turbulence fraction varies
  discontinuously in plane Couette flow, and hence that transition in the flow
  is not of directed-percolation type 
  \citep{bottin1998statistical,bottin1998discontinuous,duguet2010formation}.
  More recently, \cite{Avila_Shear_2013} conducted experiments in a counter-rotating
  circular Couette geometry (radius ratio $\eta = 0.98$) of large aspect ratio,
  and observed a variation of turbulence fraction with Reynolds
  number suggesting a continuous transition to turbulence. Further
  investigation would be needed to determine whether the transition is in the
  universality class of directed percolation.
\cite{sano2016universal} performed experiments on plane channel flow and
concluded that this flow exhibits a continuous transition to turbulence in the
universality class of directed percolation.
However, they report a critical Reynolds number (based on the centerline
velocity of the equivalent laminar flow) of $\Rey=830$, whereas other
researchers
\citep{xiong2015turbulent,duguet_private,kawahara_private,tsukahara_private}
observe sustained turbulent patches below 700. \DB{ These later authors do not
  address the question of whether the transition is continuous or
  discontinuous, and further study is needed.}

The goal of the present paper is threefold. Firstly, using a coupled-map
lattice, we present the essential issues surrounding the onset of turbulence
in a spatiotemporal setting, with particular emphasis on the case of two
space dimensions.
Secondly, we present a numerical study of a planar shear flow of unprecedented
lateral extent and show that the onset of turbulence in this flow is
continuous and is in the universality class of directed percolation.
Finally, we discuss the issue of scales in the current and past studies, and
we offer guidance to future investigations.

\section{Coupled-map lattices and directed percolation revisited}

\label{sec:CML}


Before discussing the planar shear flow, we revisit some important issues
concerning spatiotemporal intermittency and directed percolation.  The issues
can most easily be illustrated using a coupled-map-lattice (CML) model.
Such discrete-space, discrete-time models have been widely used to study the
generic behaviour arising in spatially extended chaotic dynamical systems
\cite[e.g.][]{Kaneko_Spatiotemporal_1985,Chate_Spatio_1988,
  Rolf_Directed_1998}.
Most notably they have been used as minimal models for describing the
transition to turbulence in plane Couette flow \citep{bottin1998statistical}.

\begin{figure*}
\centering\includegraphics[width=0.9\columnwidth]{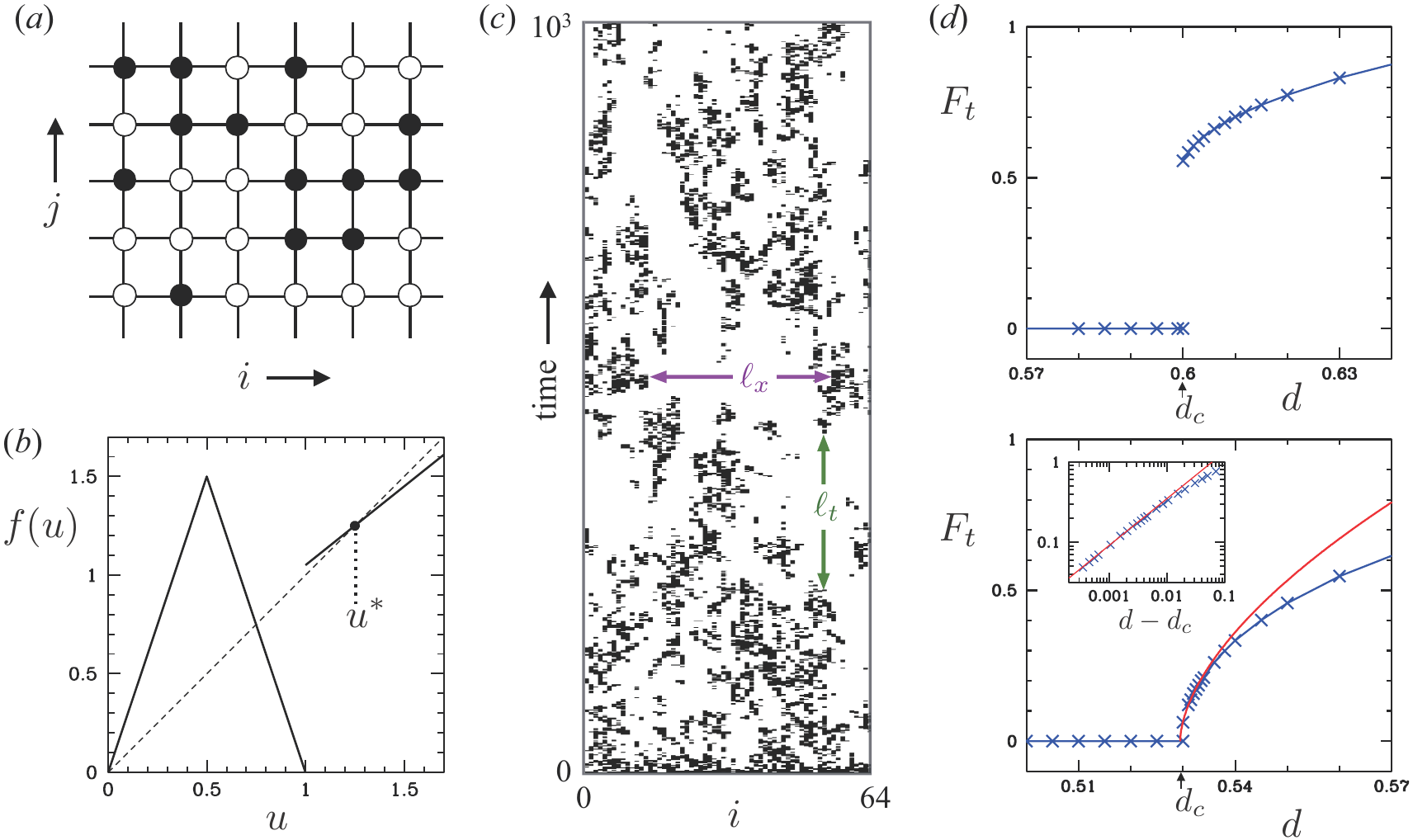}
\caption{Intermittent transition in a coupled map lattice. ($a$) Illustration
  of the lattice with nodes coloured according to whether the system is locally
  laminar (white) or turbulent (black).  ($b$) The map defining the local
  dynamics at each node. $u^*$ is a stable fixed point. ($c$) Typical time
  evolution, seen in a slice through the lattice at constant $j$, initialised
  with all sites in the turbulent state. (The spatial and temporal laminar
  gaps $\ell_x$ and $\ell_t$ are discussed in \S\ref{sec:results}.)
  ($d$) Equilibrium turbulence fraction $F_t$ as a function of the coupling
  strength $d$ in two cases: $u^*=1.25$ (top) and $u^*=1.1$ (bottom).  In the
  top case the transition to turbulence is discontinuous while in the bottom
  case it is continuous.  
  In the continuous case, close to the critical value $d_c$, $F_t$ increases
  from zero with the universal power law for directed percolation in two space
  dimensions: $F_t \sim (d-d_c)^\beta$, where $\beta \simeq
  0.583$. The red curves in the main plot and inset show this power law.
}
\label{fig:CML}
\end{figure*}

The CML model is illustrated in figure~\ref{fig:CML}. 
A state variable $u$ is defined on a discrete
square lattice (figure~\ref{fig:CML}a). Time evolution is given by discrete
updates on the lattice.
Specifically, letting $u_{ij}$ denote the state variable at point $(i,j)$, the
update rule for $u$ is
\begin{equation}
u_{ij} \leftarrow f(u_{ij}) + d \triangle_f u_{ij} 
\label{eq:CML}
\end{equation}
where the first term, $f(u_{ij})$, is the local dynamics and the second term
is a nearest-neighbour diffusive-like coupling.  
\DB{
Lattice sites are updated asynchronously by cycling through $(i,j)$ in a
random order for each step, as described by \cite{Rolf_Directed_1998}.}
The control parameter is the coupling strength $d$.

The local dynamics are given by the map $f$ shown in
figure~\ref{fig:CML}(b). It has a ``turbulent'' tent region for $0 \le u \le
1$ and a ``laminar'' region $u>1$ surrounding the stable fixed point at
$u^*$.  In the absence of coupling, a turbulent site evolves
chaotically and eventually makes a transition to the laminar state.
\DB{Once a site becomes laminar, it will remain so indefinitely.  The
  laminar fixed point is referred to as an absorbing state.}
Hence, the local dynamics are a simple caricature of a subcritical shear flow
with coexisting turbulent and laminar flow states.
Because the model is only a slight generalisation of those appearing
in numerous past studies, we relegate the details to the appendix.

We are primarily interested in the long-time dynamics of the system. 
We start from an initial condition with randomly selected values
  within the turbulent region.
%
%
Figure~\ref{fig:CML}(c)
shows the evolution as seen in a one-dimensional slice through the lattice.
The main quantity of interest is the turbulence fraction $F_t$, which is the
fraction of sites in the turbulent state.
After some time, the system will reach a statistical equilibrium and we can
obtain the equilibrium value of $F_t$.
If this is zero, then the system is everywhere in the laminar (absorbing)
state. If it is non-zero, then at least some turbulence persists indefinitely.

A basic question is: how does the turbulence fraction at equilibrium depend on
the coupling strength $d$, and in particular, how does it go from zero to non-zero?
Figure~\ref{fig:CML}(d) shows the two distinct cases: one discontinuous and
one continuous.  In the discontinuous case, there is a gap in the possible
values of $F_t$. Long-lived transients with small turbulence fraction can be
observed for values of $d$ below $d_c$, the critical value of $d$, but the 
system simply cannot indefinitely maintain a small level of
turbulence, no matter how large the system size.
On the other hand, in the continuous case, $F_t$ becomes arbitrarily small
(in the limit of infinite system size)
as $d$ approaches $d_c$ from above.  In this case the system behaves in
accordance with the power laws of directed percolation.
In particular, as is shown, the turbulence fraction grows as $F_t \sim (d -
d_c)^\beta$, where $\beta \simeq 0.583$; see \cite{Lubeck_universal_2004}. We
will discuss the other important power laws later when we analyse the planar
fluid flow.

Note that on any finite lattice the minimum possible
  non-zero turbulence fraction is 1/$K$, (i.e.\ just one turbulent site),
  where $K$ is the total number of lattice points. This means that even if the
  transition is continuous in principle, some discontinuity in the turbulence
  fraction from finite-size effects will be present in any numerical study.
  It is by investigating scaling behaviour, such as the log-log plot in
  figure~\ref{fig:CML}(d), that one gains confidence in the nature of the
  transition.

\subsection{Connection to turbulent transition and directed percolation}

The difference between the continuous and discontinuous cases presented in
figure~\ref{fig:CML}(d) is only the location of the laminar fixed point $u^*$ in the map $f$.
Hence, either case could in principle correspond to a
shear flow and there is no way to know {\em a priori} what type of transition
could be expected.
This point was well understood by the Saclay group in their early studies on
transition in plane Couette flow
\cite[e.g.][]{bottin1998statistical,bottin1998discontinuous,berge1998espace,
  Manneville_Transition_2016}.
Those experiments suggested a discontinuous transition to turbulence, based
not only on the turbulence fraction, but also on the nature of transients
below the critical point.
Although we will argue that the physical size of those experiments was
too small to produce a continuous transition, the conclusion reached was
reasonable at the time.

More generally, directed percolation describes a {\em stochastic process}
involving active and absorbing states (or equivalently bonds between sites
that are randomly open or closed).  
As \citet[][Section 4.2]{Manneville_Transition_2016} notes, deterministic
iterations of continuous variables coupled by diffusion will not necessarily
behave in the same way as the directed percolation process.
The CML model presented here demonstrates this point. Depending on parameters,
the system might, or might not, show a continuous transition in the
universality class of directed percolation.
Notwithstanding Pomeau's conjecture, it is even less immediately evident
  that the full Navier--Stokes equations will behave in the same way, owing to
  the global nature of the pressure field for example.
(For more technical details on absorbing state transitions, we refer the
reader to \cite{Lubeck_universal_2004}, and references therein. One can find
there details of the Janssen-Grassberger conjecture,
\citep{janssen1981nonequilibrium,grassberger1982phase}, concerning
the ubiquity of the directed percolation universality class.)

From hereon we shall use the notation of directed percolation and refer to the
case of two space dimensions as (2+1)-D, meaning two spatial and one temporal
dimension.  In this notation, the spatial dimensions are referred to as
perpendicular ($\perp$) and the temporal dimension as parallel ($\parallel$).

\section{Waleffe flow}
\label{sec:Waleffe}

Pinning down the details of transition requires very large system sizes.  For
example, in the quasi-one-dimensional experiments of
\cite{lemoult2016directed}, the long direction was more than $2700$ times the
fluid gap.
Our goal is to achieve something approaching this size, but in two spatial
directions and in a computational framework.
To this end, we shall study a cousin of Couette flow, commonly referred to
as Waleffe flow. This is the shear flow between parallel stress-free
boundaries, driven by a sinusoidal body force.
The two related computational advantages of this flow are that it lacks high-shear
boundary layers near the walls 
and that the wall-normal dependence of the flow can be accurately represented by
a few trigonometric functions. 
%
As shown in \cite{chantryTB}, a poloidal-toroidal representation with at most
four trigonometric modes in the wall-normal direction, $y$, is capable of
capturing turbulent bands and spots, the building blocks of turbulent-laminar
intermittency.  A Fourier representation is used for the large streamwise,
$x$, and spanwise, $z$, directions.

In \cite{chantryTB}, we showed that Waleffe flow corresponds closely
to the interior of plane Couette flow, leading to a change in length
scales from $2h$ (the gap between walls in plane Couette flow) to $1.25h$
(the Couette interior region) for Waleffe flow. 
{Furthermore, this argument regarding the interior region leads to a comparable velocity
  scale $U = 1.6V$, with $V$ the maximum velocity of laminar Waleffe flow.
  The Reynolds number of the flow is then $\Rey = U h/\nu$, where $\nu$ is the
  kinematic viscosity.  }
%
The sole change from \cite{chantryTB} is the addition of a small
horizontal drag force $-\sigma (u \mathbf{e}_x + w \mathbf{e}_z)$ to
the Navier-Stokes equation.
Such a term, usually called Rayleigh or Ekman friction, is used in many
hydrodynamic modelling contexts to approximate the effect of friction
due to a solid boundary that has been omitted from the model. 
In geophysics \citep[chap. 4]{marcus1998model,pedlosky2012geophysical}
the inclusion of this term is the standard method of including
the first-order departure from geostrophic flow due to the Ekman boundary layer
between a stationary bottom and a rotating bulk.
In their study of electromagnetically driven Kolmogorov flow in an electrolyte, 
\cite{suri2014velocity} include such a term in their depth-averaged model
of an assumed Poiseuille-like profile 
in order to account for the presence in their experiment of
a solid boundary at the bottom of the fluid layer.

In our case, we introduce this force in order to damp flows with no
curvature in $y$ and very little curvature in $x$ and $z$, which decay
extremely slowly in Waleffe flow and which are not present at all in
Couette flow.  Our purpose is to use Waleffe flow to mimic the bulk
region of Couette flow, which it does very well except for this point.
The value $\sigma=10^{-2}$ reproduces the damping to which these modes
would be subjected in the wall regions of the corresponding Couette
flow. In very large domains, without this damping the recovery of the
laminar flow after spot decay is very slow. Beyond
this, the damping has no effects on the phenomenology of Waleffe flow.

\begin{figure*}
\centering\includegraphics[width=1.0\columnwidth]{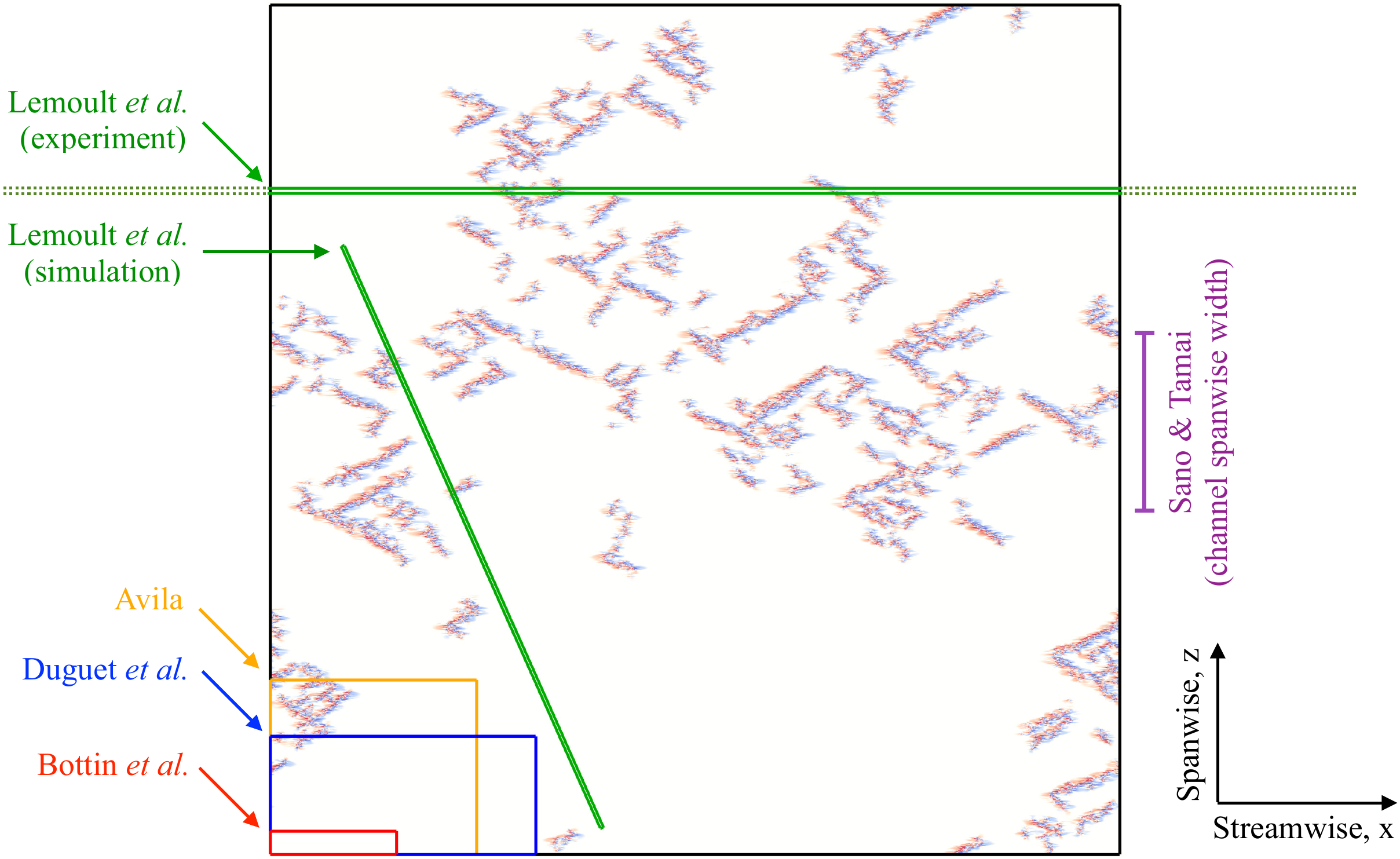}
\caption{Intermittent turbulence typical of that found slightly above the
  onset of sustained turbulence. Visualized is streamwise velocity in the
  midplane {at \Rey = 173.824 after $1.2\times 10^6$ time units}.  Laminar 
  flow is seen as white.  The streamwise and spanwise size
  of the computational domain is $2560h \times 2560h$.
The turbulence fraction is $F_t \approx 0.1$ and 
the reduced Reynolds number is $\epsilon = (Re - Re_c)/Re_c = 1.4 \times
 10^{-4}$.
%
%
For reference, the Couette domains of \cite{bottin1998discontinuous},
\cite{duguet2010formation}, {\cite{Avila_Shear_2013}} and
\cite{lemoult2016directed} are overlaid in red, blue, {orange} and
green respectively. The full streamwise length of the
\cite{lemoult2016directed} experiment exceeds the figure size and is
not fully shown.  The spanwise width of the \cite{sano2016universal}
channel experiment is indicated in purple on the right.  }
\label{fig:Domain}
\end{figure*}

%
%
%

We shall present results for domains of size $[1280h,1.25h,1280h]$,
$[2560h,1.25h,2560h]$ and $[5120h,1.25h,1280h]$. 
Our largest square domain is plotted in figure~\ref{fig:Domain}, where we show
a representative turbulent state slightly above the onset of turbulence.
%
%
For this domain the highest resolved wavenumber in each horizontal direction
is 2047, with 3/2 dealiasing used (leading to a grid spacing of 0.42). The same turbulence
fractions are found in simulations with twice the resolution.

For context, the experiments of \cite{bottin1998discontinuous} used a domain
of size $[380h,2h,70h]$, \cite{prigent2003long} used a domain of size
$[770h,2h,340h]$ and \cite{Avila_Shear_2013} used a domain of size
$[622h,2h,526h]$.
To date the largest simulations have been those of
\cite{duguet2010formation}, who considered a domain of size
$[800h,2h,356h]$. Both \cite{bottin1998discontinuous} and
\cite{duguet2010formation} report evidence of a discontinuous transition,
unable to sustain turbulence fractions significantly below 0.4,
while \cite{Avila_Shear_2013} observed evidence of a continuous
transition, with sustained turbulence fractions as small as $0.07$.

The flow at $(x,z,t)$ is defined as turbulent if $E(x,z,t) > E_T$, where
$E(x,z,t)$ is the $y$-integrated energy of the velocity deviation from the
laminar state and $E_T=0.01$ is a threshold.
Varying $E_T$ between 0.001 and 0.05 changes only slightly the size of
patches deemed turbulent and has no effect on any of the scaling
relationships to follow.

\section{Results}
\label{sec:results}

\begin{figure}\begin{center}
\includegraphics[width=0.48\columnwidth]{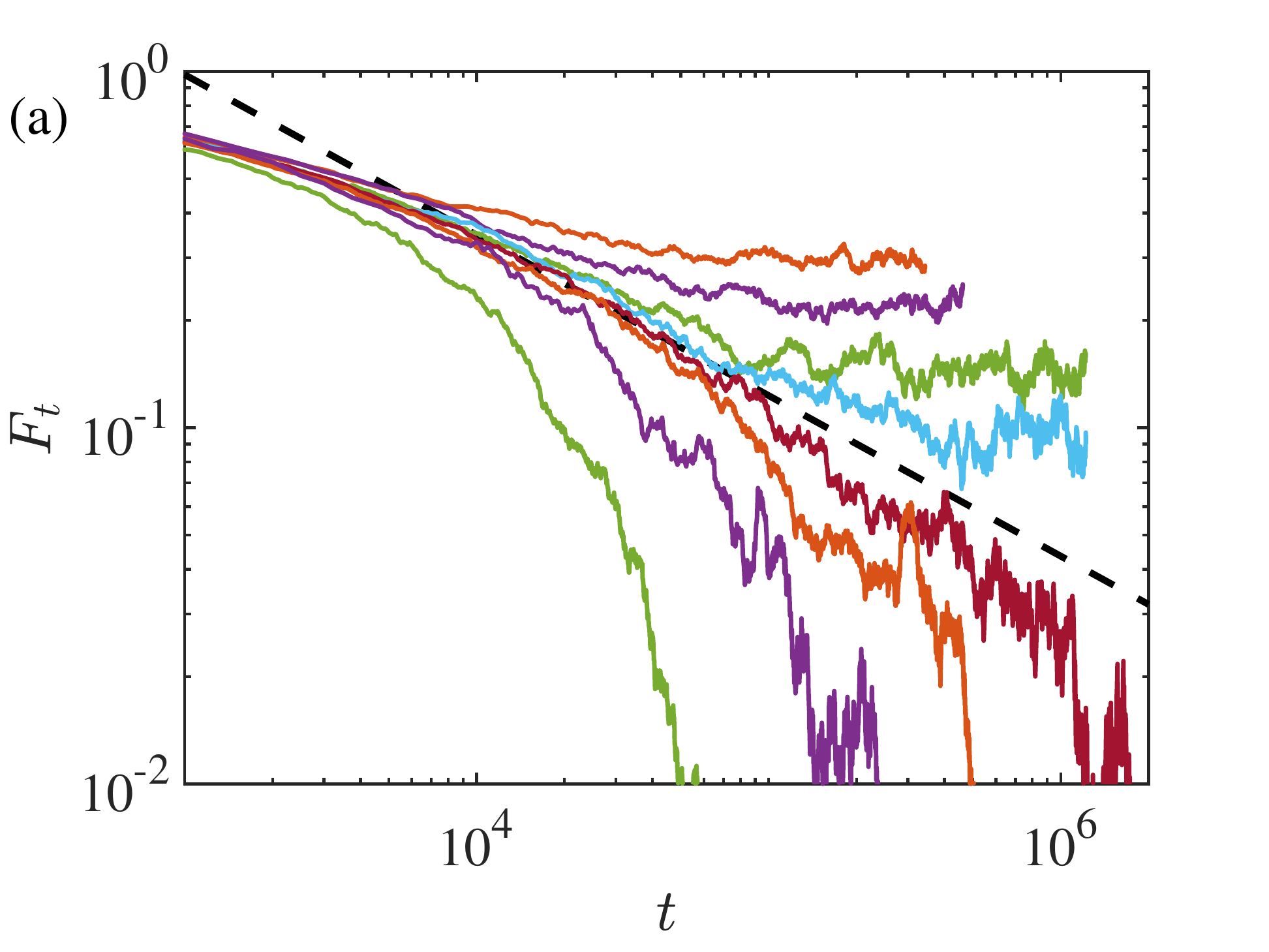}
\includegraphics[width=0.48\columnwidth]{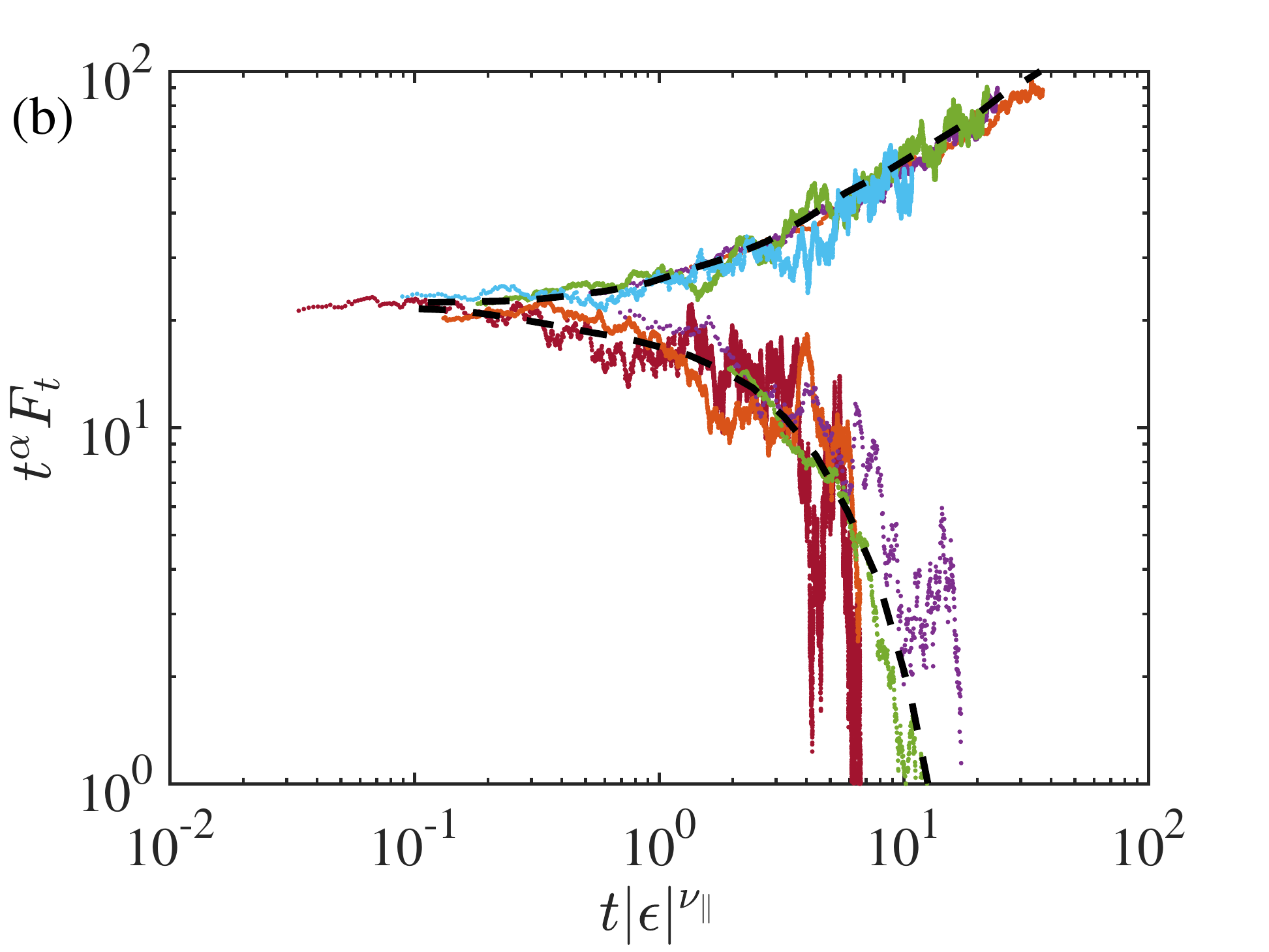}
\end{center}
\caption{(a) Turbulence fraction as function of time for a range of
  Reynolds numbers with an initial condition of uniform turbulence. Above
  criticality, the turbulence fraction saturates at a finite value,
  and below it falls to zero. At criticality, the turbulence fraction
  decays in time as a power law $F_t\sim t^{-\alpha}$ with the (2+1)D
  directed percolation exponent $\alpha \simeq 0.4505$ (dashed
  line). Coloured lines, for decreasing turbulence fractions
  correspond to Reynolds numbers
  $[173.952 ,173.888,173.840,173.824,173.792, 173.773, 173.696,
  173.568]$.
  (b) Data above and below criticality collapse onto two scalings
  (black dashed curves) when the directed percolation exponents are
  used to rescale time and turbulence fraction. }
\label{fig:Decay}
\end{figure}

In figure~\ref{fig:Decay}(a), we plot the time evolution of the
turbulence fraction $F_t$ for a series of Reynolds numbers.  Each run
was initialised from uniform turbulence and run until a saturated
turbulence fraction was reached \citep[quench protocol,
see][]{bottin1998statistical}.
Below a critical value $\Rec=173.80$ 
(to five significant figures), the turbulence fraction eventually
falls off to zero, while above $\Rec$ it saturates at a finite value.
($\Rec$ for this system differs from that of plane Couette flow).

\begin{figure}
\centering{\includegraphics[width=.36\columnwidth]{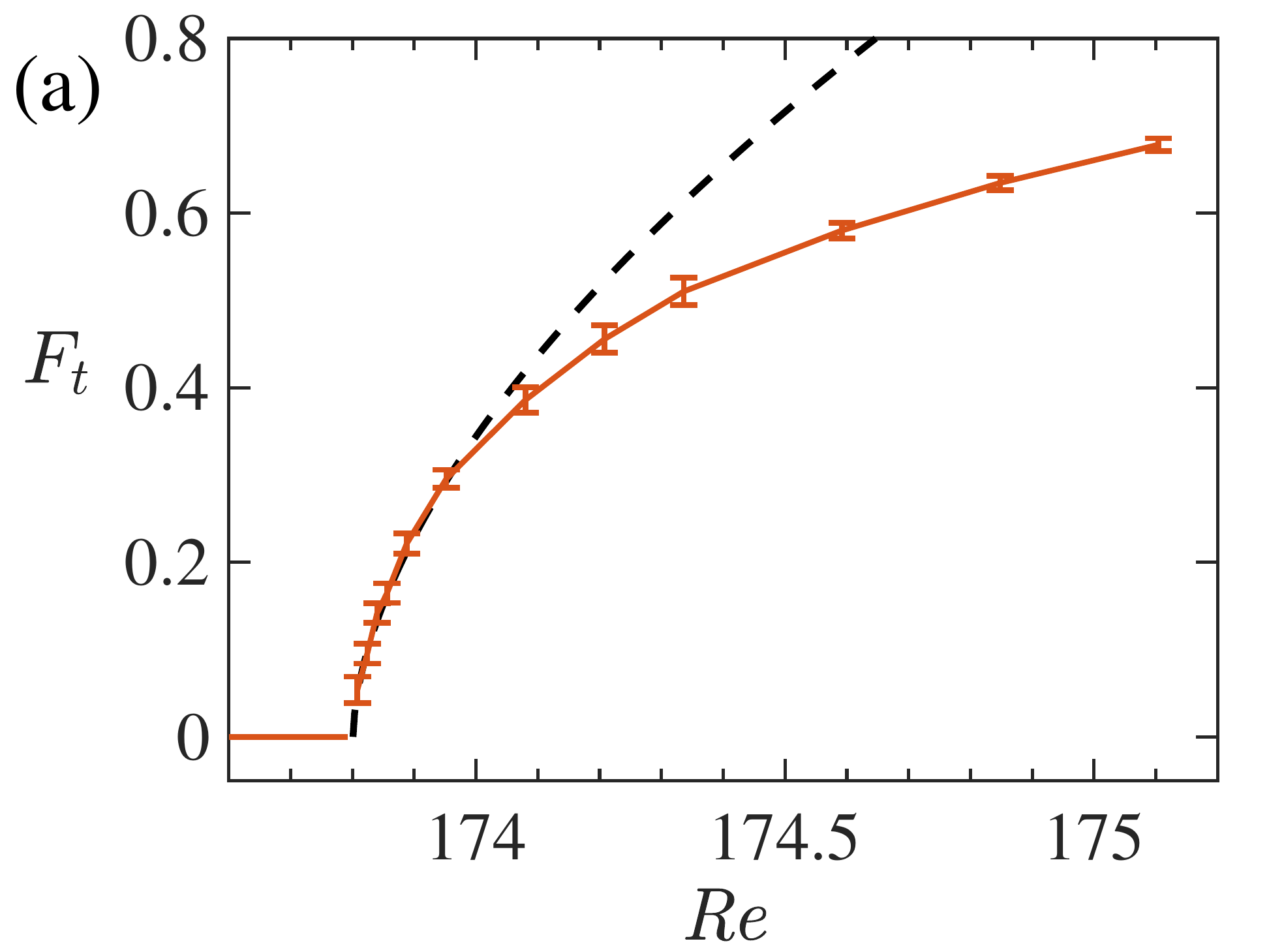}
\includegraphics[trim={0.4cm 0cm 5.8cm 0cm},clip,width=.25\columnwidth]{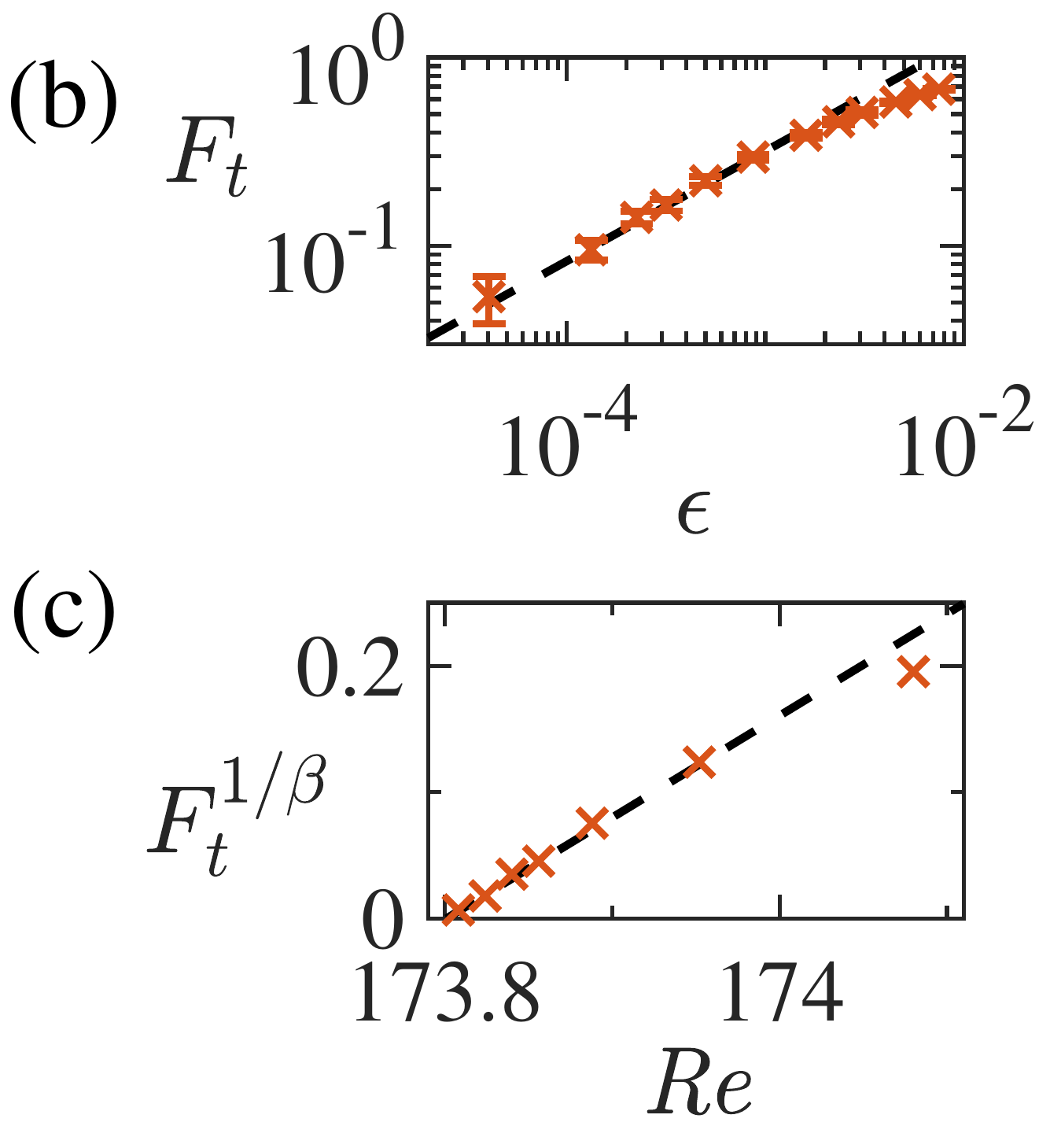}
\includegraphics[trim={0cm 0cm 0.25cm 0cm},clip,width=.36\columnwidth]{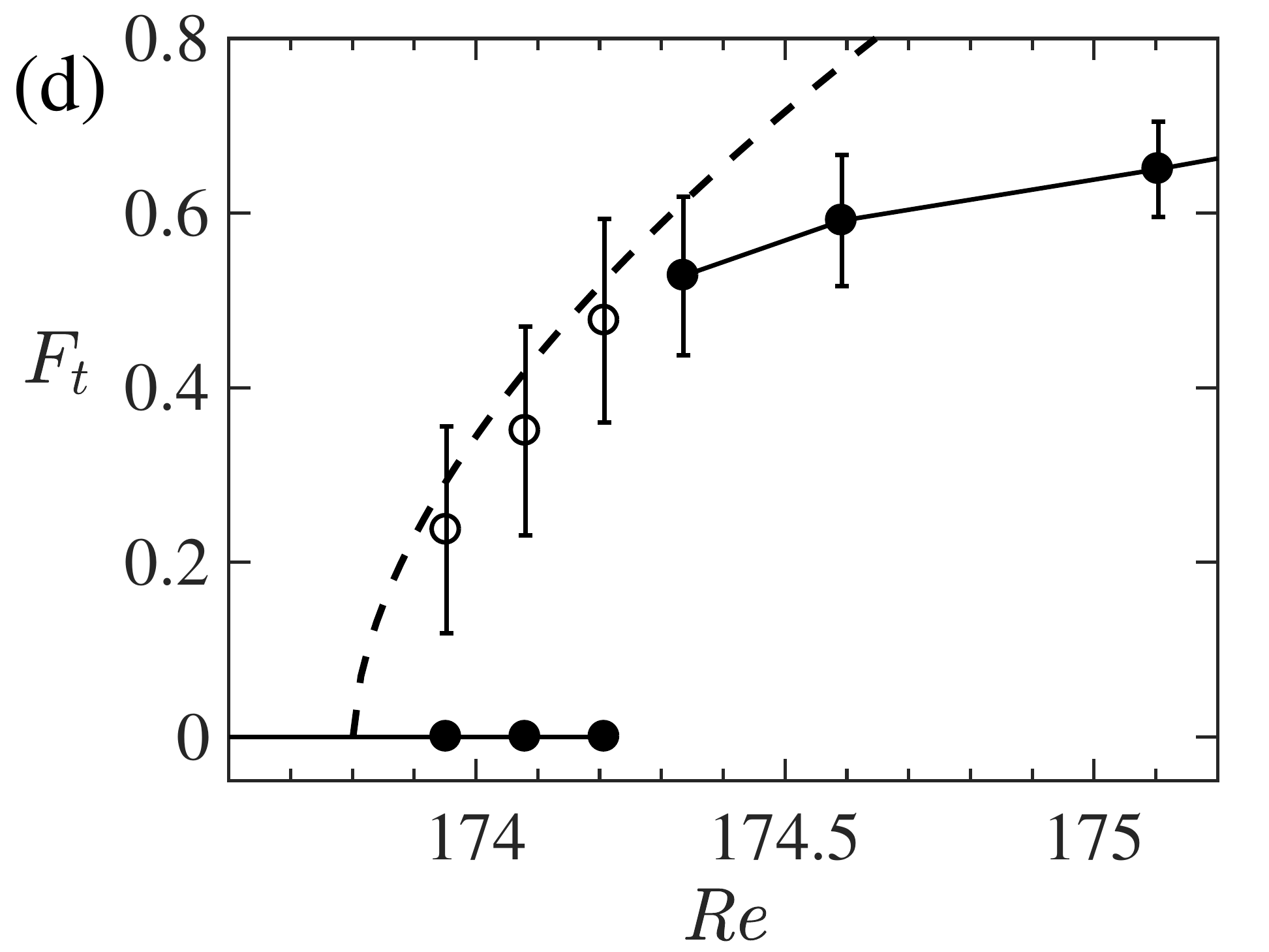}}
\caption{Bifurcation diagrams for the transition to turbulence.  (a)
  Continuous transition in a large domain: $[2560h,1.25h, 2560h]$. Equilibrium
  turbulence fraction $F_t$ is plotted as a function of $\Rey$. Points and
  error bars denote mean and standard deviation of $F_t$. Black dashed curved
  shows the directed percolation power law. {(b)} Log-log plot of
  the same data in terms of $\epsilon = \left(\Rey-\Rec\right)/{\Rec}$, where
  $\Rec = 173.80$.  Near criticality the data is consistent with $F_t \sim
  \epsilon^\beta$ with $\beta \simeq 0.583$. 
   (c) $F^{1/\beta}$ against $\Rey$ showing linear behaviour. 
  {(d)} Discontinuous transition in a domain of size $[380h,1.25h,70h]$,
  approximately that of the experiments by \cite{bottin1998statistical}.  (See
  figure~\ref{fig:Domain}.)  Filled points denote sustained turbulence, while
  open points denote the turbulence fraction of long-lived transient
  turbulence.
  The dashed curve is the directed-percolation power law from the large
  domain.}
\label{fig:Beta}
\end{figure}

In figure \ref{fig:Beta}(a) we plot the equilibrium turbulence fraction as a
function of $\Rey$.  We find clear evidence for a continuous transition in
which small $F_t$ can be sustained given a sufficiently large domain.  The
saturated turbulence fraction follows a power law $F_t\sim \epsilon^\beta$ where
$\epsilon\equiv(\Rey-\Rec)/\Rec$. 
$\Rec$ is determined as the value of $\Rey$ that minimises the mean squared 
error of a linear fit of the logarithms of $\epsilon$ and $F_t$ (see figure \ref{fig:Beta}b). 
This linear fit
estimates $\beta=0.58\pm 0.04$ with a $95\%$ confidence interval. This agrees with 
the (2+1)-D directed percolation value
$\beta \simeq 0.583$ (dashed line).
Figure~\ref{fig:Beta}(d) shows the turbulence fraction obtained 
from our system in a domain
whose size is that of the experiments of \cite{bottin1998statistical}.
(See also figure~\ref{fig:Domain}.)
As was observed experimentally, below a turbulence fraction of about $0.5$,
turbulence appears only as a long-lived transient state, and hence 
the equilibrium turbulence fraction exhibits a discontinuous transition. 
This strongly suggests that the discontinuous transitions reported for plane
Couette flow
\citep{bottin1998statistical,bottin1998discontinuous,duguet2010formation} are
due to finite-size effects.
Interestingly, long-lived transient states in the small system have turbulence
fractions close to those for equilibrium states in our large domain; they just
are not sustained states. Also note that the critical Reynolds number is not greatly
affected by system size.

\begin{figure*}
\includegraphics[width=0.325\columnwidth]{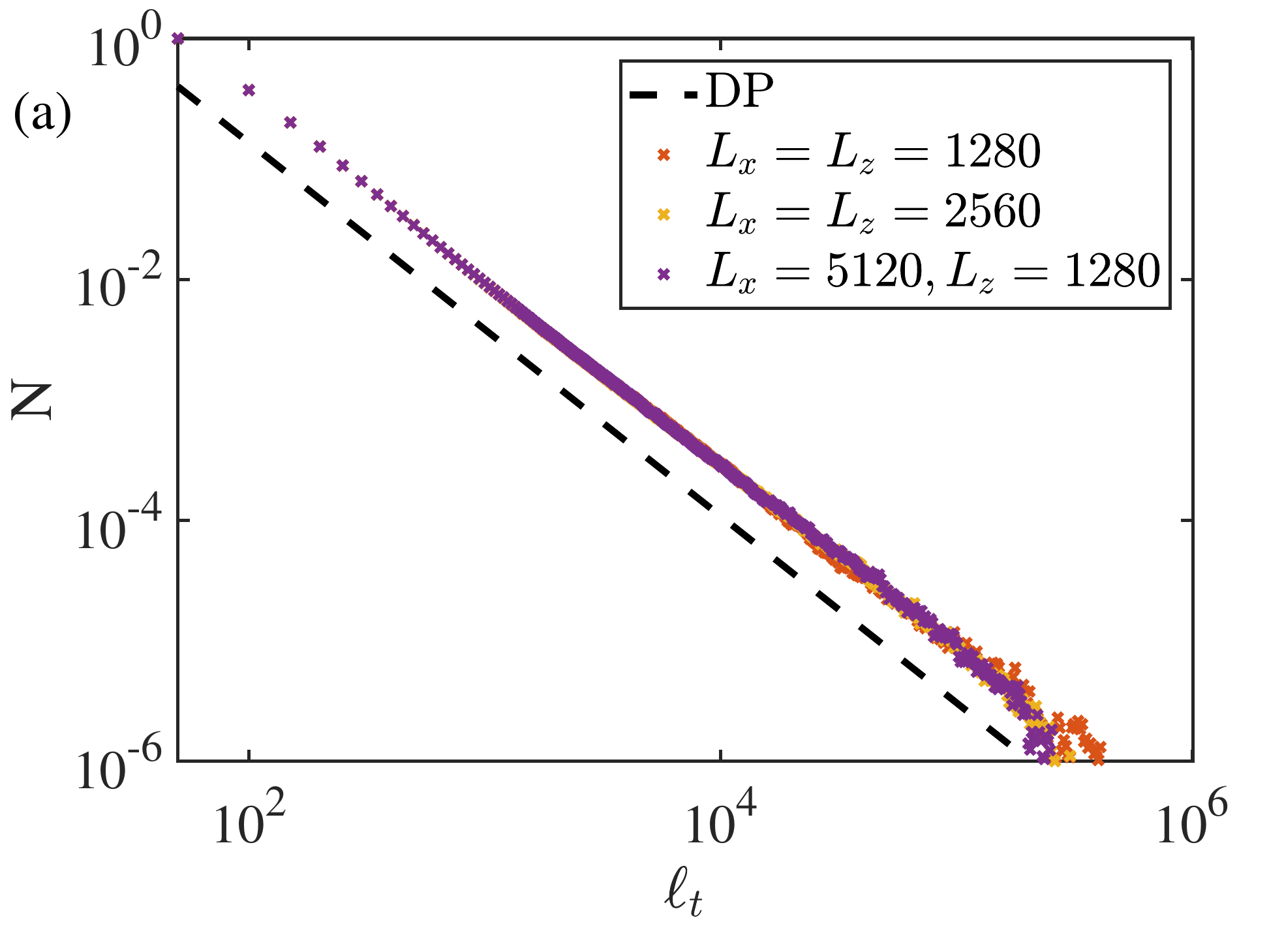}
\includegraphics[width=0.325\columnwidth]{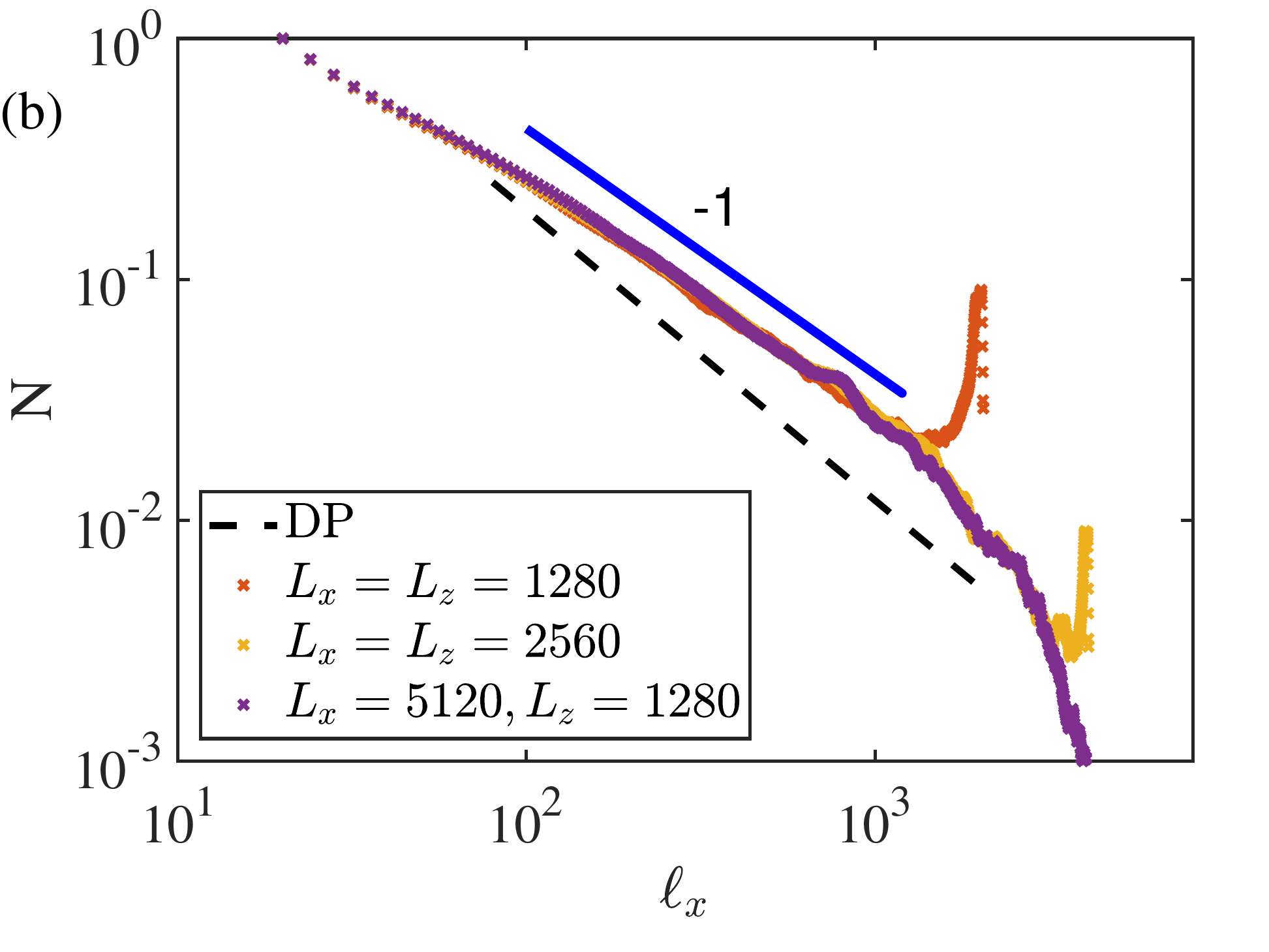}
\includegraphics[width=0.325\columnwidth]{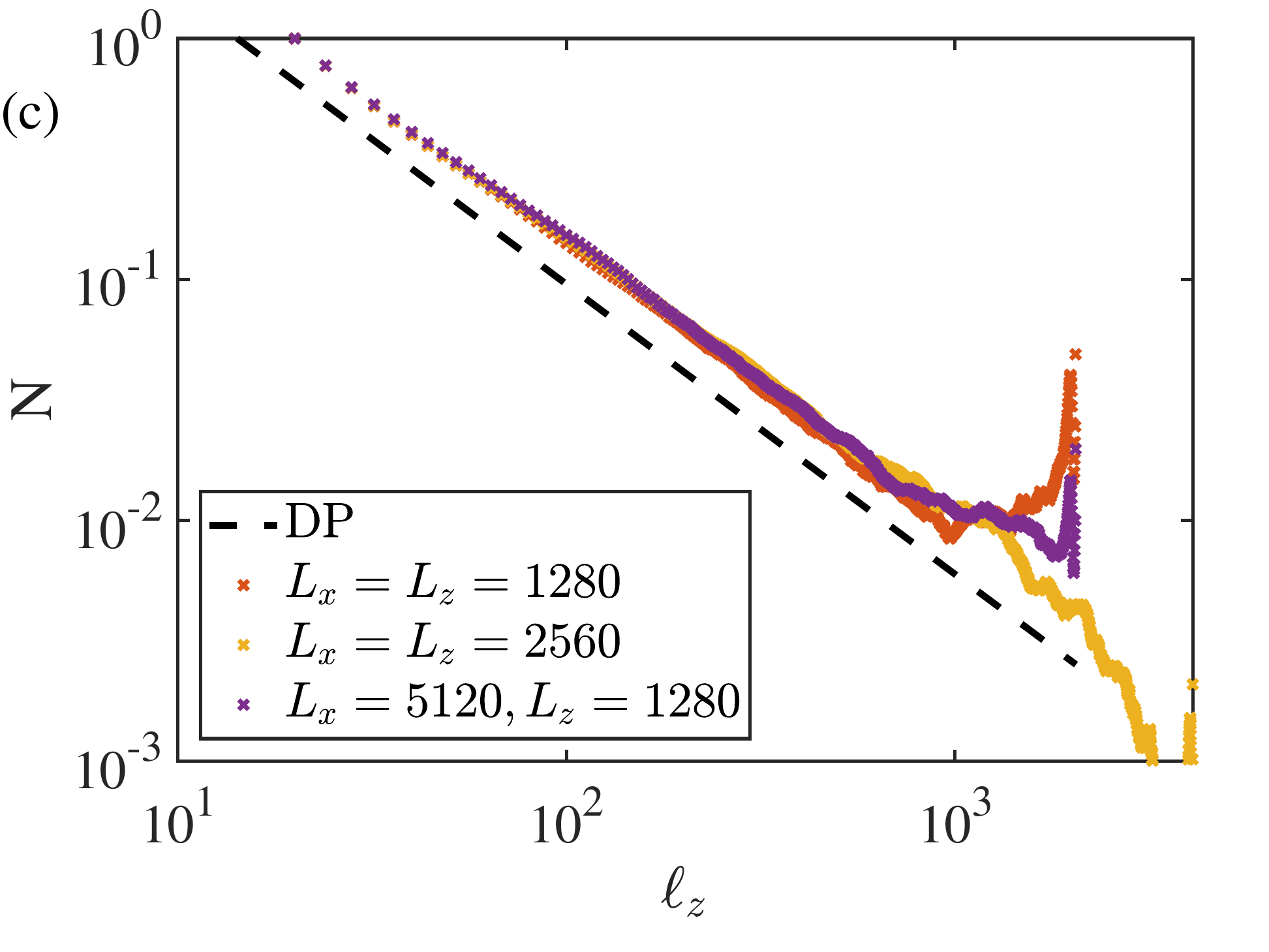}
\includegraphics[width=0.325\columnwidth]{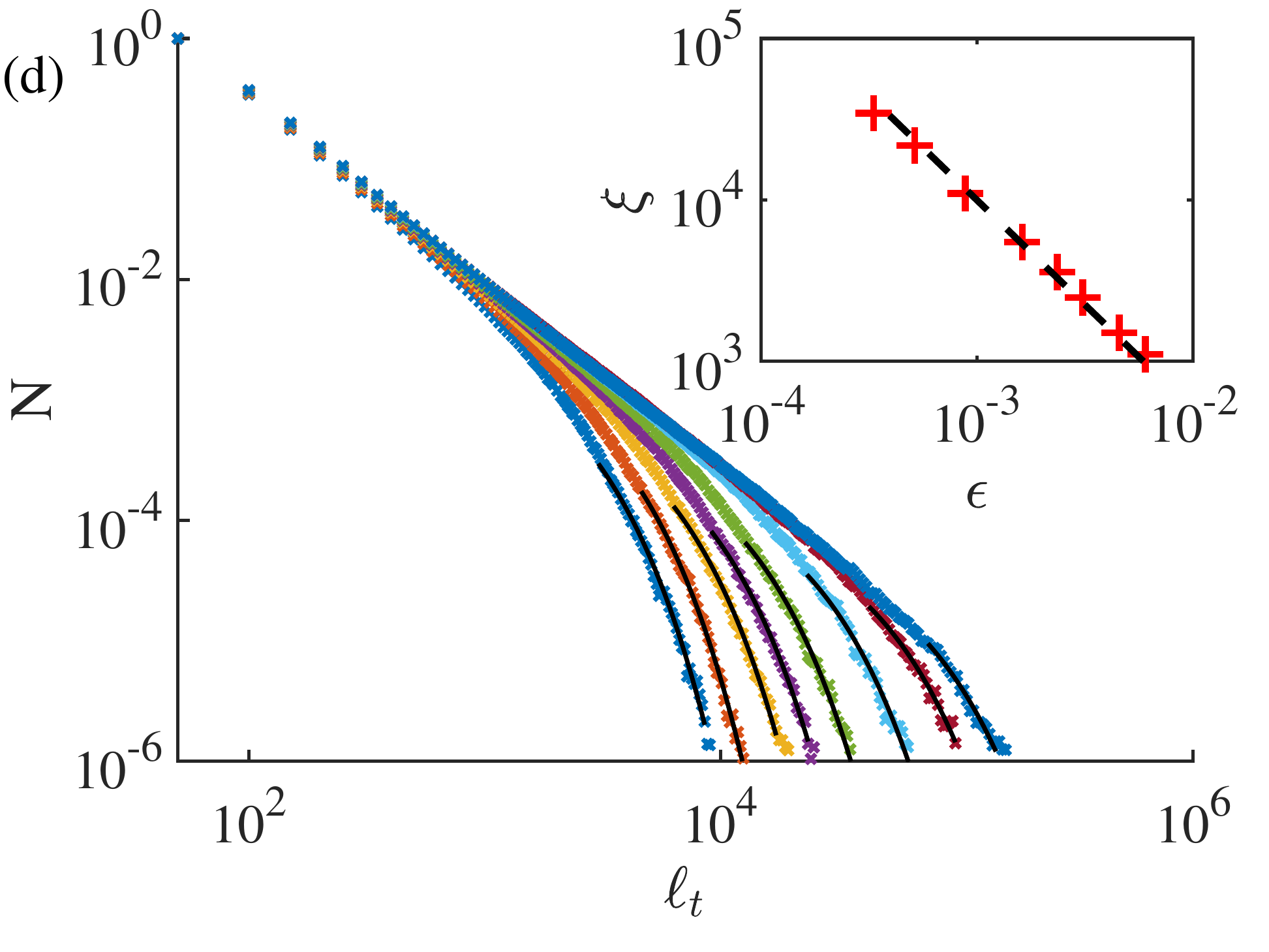}
\includegraphics[width=0.325\columnwidth]{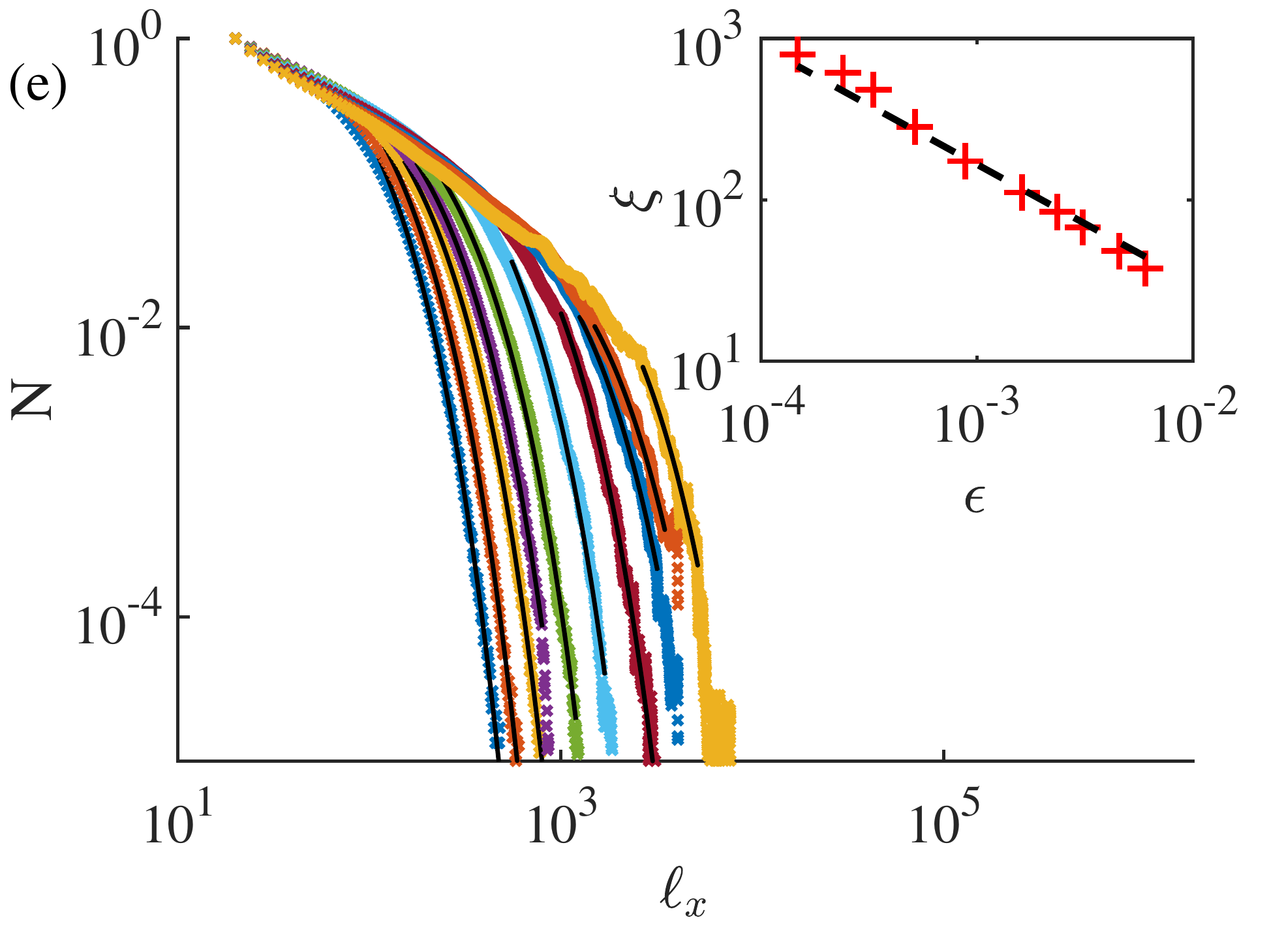}
\includegraphics[width=0.325\columnwidth]{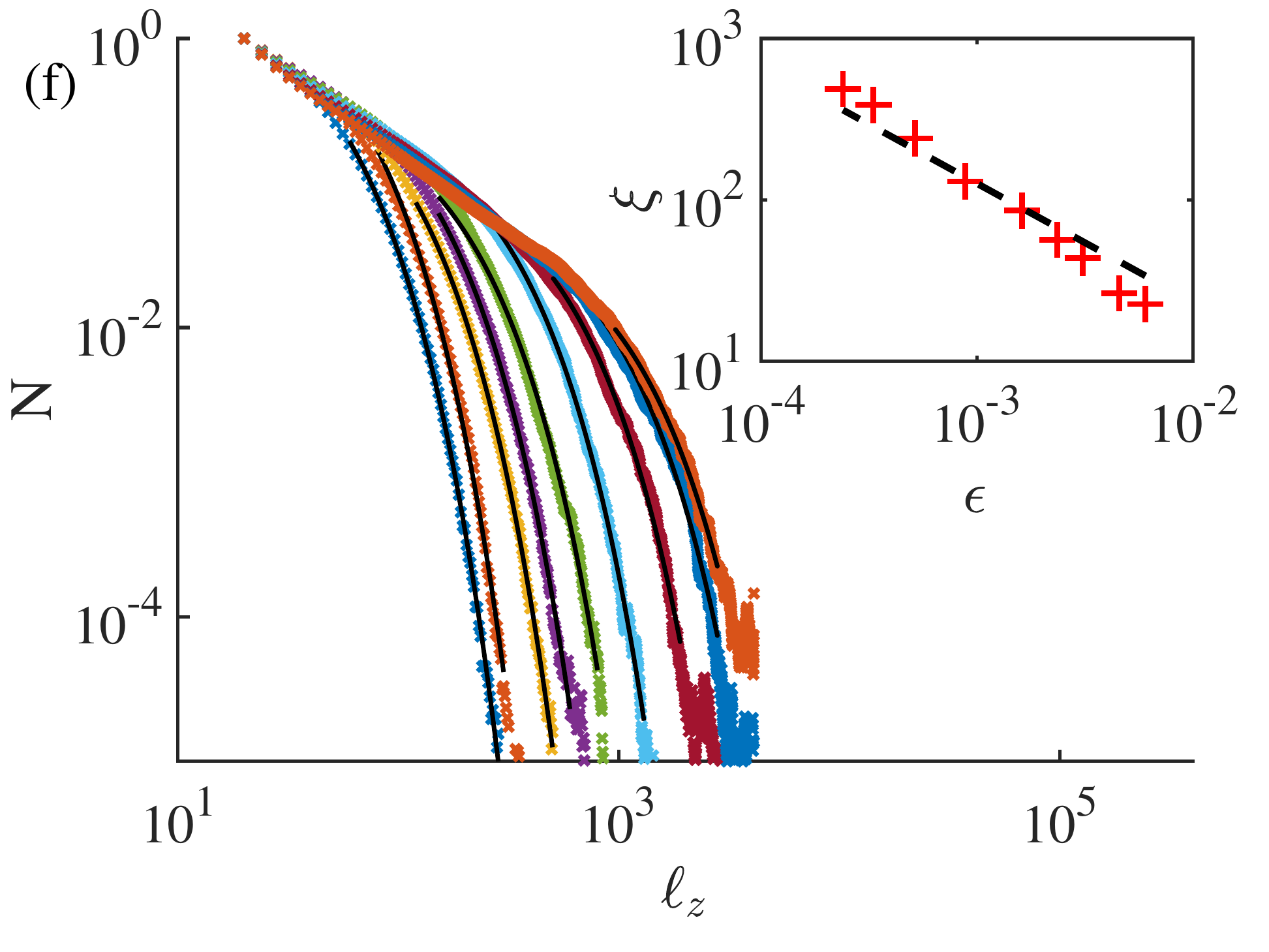}
\includegraphics[width=0.325\columnwidth]{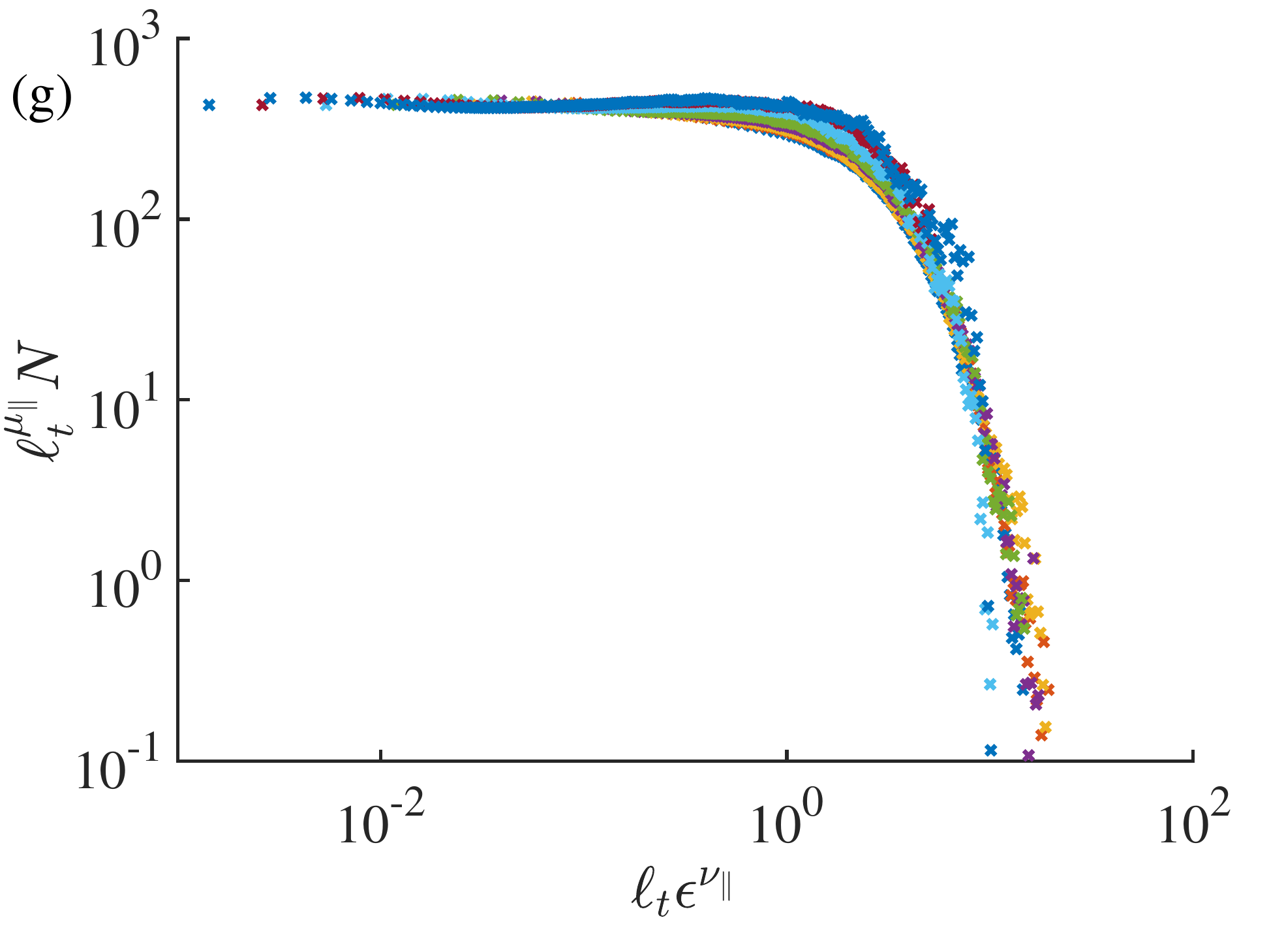}
\includegraphics[width=0.325\columnwidth]{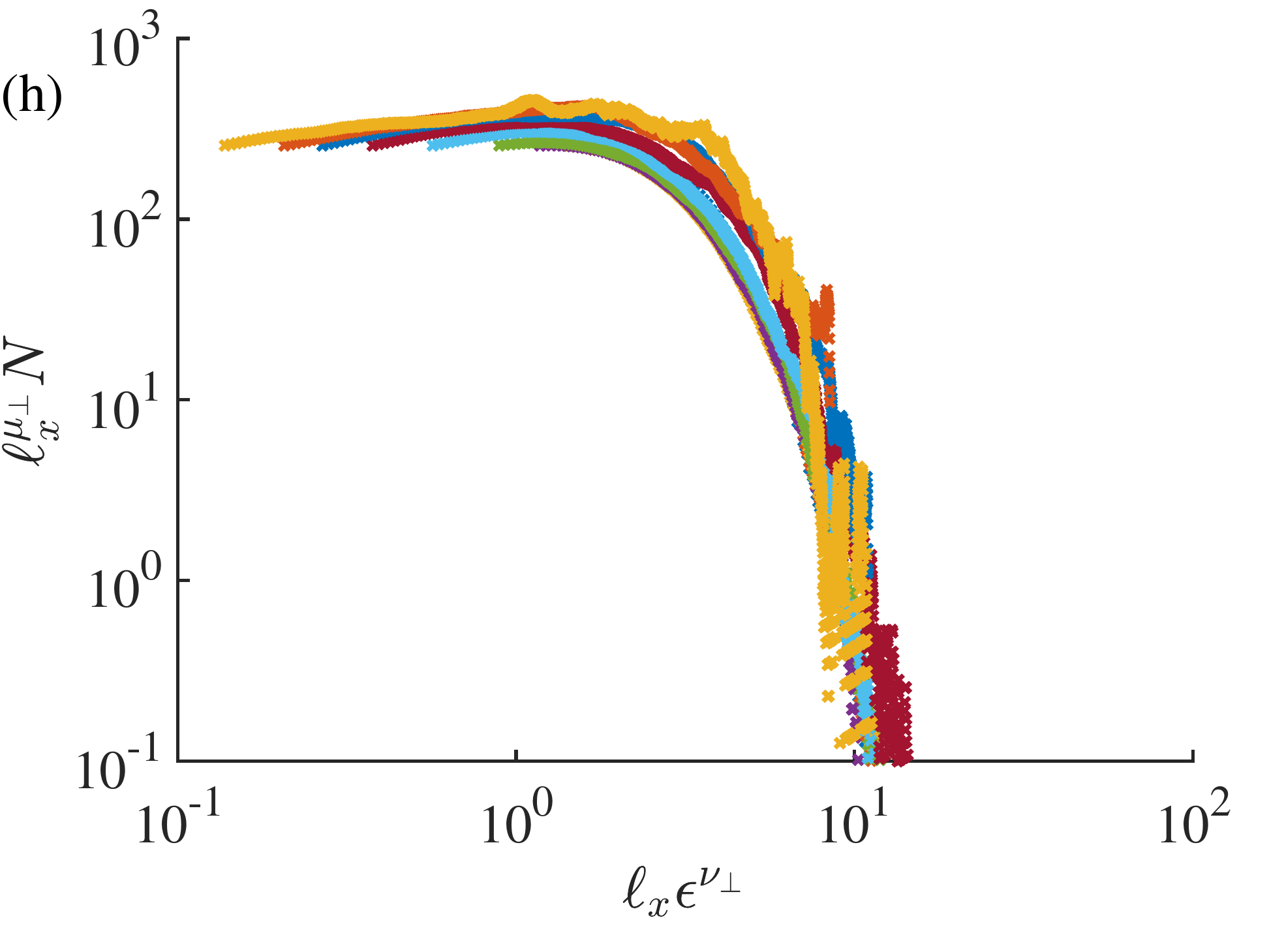}
\includegraphics[width=0.325\columnwidth]{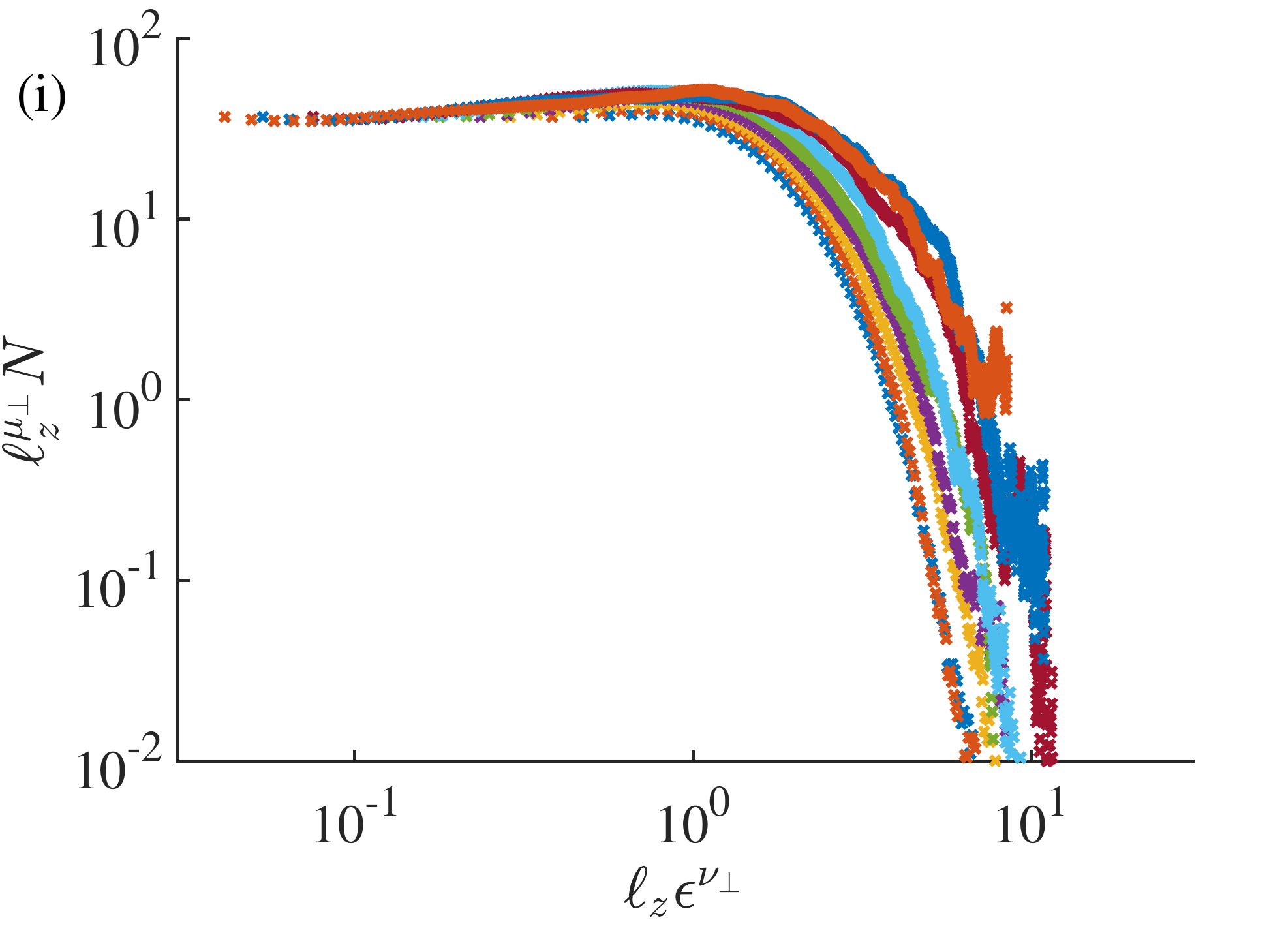}
\caption{Exponents for temporal and spatial correlations. Top row shows
distributions of laminar gaps in (a) time and the two spatial directions,
(b) $x$ and (c) $z$, for a variety of domain sizes at $\Rey=173.824$
($\epsilon =1.4 \times 10^{-4}$).   {$N$ is the gap count,
normalised by the shortest gap count}. The directed percolation scalings
($\mu_\parallel \simeq 1.5495 $ and $\mu_\perp \simeq 1.204$) are plotted as dashed
lines and show excellent agreement with the $t$ and $z$ gaps.  For the
$x$-gaps a power law closer to $-1$ is observed.
Middle row, (d-f), shows the exponential tails of the gap distributions for
a several values of $\epsilon$ just above criticality. Increasing domain
sizes are used for points closer to criticality. Insets show scaling of
correlation lengths $\xi$ with $\epsilon$ together with the directed
percolation exponents: $\nu_\parallel \simeq 1.295$ and $\nu_\perp \simeq 0.733$.
Bottom row, (g-i), collapse of the data using directed percolation
power laws $\mu$ and $\nu$ (see axis labels/text).}
\label{fig:LamG}
\end{figure*}

To substantiate whether a system is in the directed percolation universality
class, it is necessary to verify three independent power-law scalings close to
criticality; see \cite{takeuchi2009experimental} whose approach we will follow
closely.  
Having demonstrated the scaling of $F_t$ (exponent $\beta$),
we now turn to scalings associated with temporal and spatial correlations.

One approach to determining the correlations is via the distribution of
laminar gaps at $\Rey \simeq \Rec$, (top row of figure~\ref{fig:LamG}).
The flow has a temporal laminar gap of length $\ell_t$ if $E(x,z,t) > E_T$ and
$E(x,z,t+\ell_t) > E_T$ but $E(x,z,t^\prime) < E_T$ for $0<t^\prime<\ell_t$.
Spatial gaps $\ell_x$ and $\ell_z$ are defined similarly.  Such gaps are
illustrated for the CML model in figure~\ref{fig:CML}(c).  From simulations
just above $\Rec$, we generate gap distributions by measuring and binning the
laminar gaps within the intermittent flow  {once the turbulence fraction 
has saturated}.  Given the anisotropy between the
streamwise and spanwise directions, we measure gaps in these directions
separately.
At criticality, a system within the directed-percolation universality class
displays power-law behaviour, $N\sim \ell^{-\mu}$,
where $N$ is the number of gaps of length $\ell$.
The temporal gaps, figure~\ref{fig:LamG}(a), show excellent scaling 
with the directed-percolation temporal exponent $\mu_\parallel\simeq 1.5495$
This is also true of the spanwise gaps, figure~\ref{fig:LamG}(c), 
with the spatial exponent $\mu_\perp\simeq 1.204$. 
However,
the streamwise laminar gaps, figure \ref{fig:LamG}(b), do not show a clear,
extended power law, and to the extent that there is a power law, the exponent is
closer to 1 than to $\mu_\perp \simeq 1.204$.
Indeed, \cite{takeuchi2007directed} observed that the laminar gap
distribution in one direction of a liquid crystal layer
had an exponent closer to 1 than to $\mu_\perp \simeq 1.204$,
This is also true for our simulations of a non-isotropic CML very slightly 
above the critical point,
thus indicating that the issue here is that the flow is not exactly at
$\Rec$, as it should be for the scaling to hold.
Although the gap distribution in each spatial direction should show
power-law behaviour, these may converge at different rates as $\Rey \rightarrow \Rec$.

Given the poor agreement for the streamwise gaps, we use a second approach to
measure the percolation exponents which does not rely on simulations at $\Rec$.
Away from criticality, power-law behaviour will be seen only over a finite range
of temporal and spatial gap lengths. Beyond these
lengths, exponential tails are expected of the form $N \sim \exp(-\ell/\xi)$,
with correlation lengths $\xi$ diverging as $\epsilon$ goes to zero: $\xi \sim
\epsilon^{-\nu}$.
In the  {middle} row of figure~\ref{fig:LamG} we fit exponential tails for
several values of $\epsilon$. In the insets we plot $\xi$ as a function of
$\epsilon$ and compare with the expected exponents for directed percolation.
The exponents $\mu$ and $\nu$ are exactly related via
$\mu = 2 - \beta / \nu$,
thus giving $\nu_\parallel \simeq 1.295$ and
$\nu_\perp \simeq 0.733$ \citep{Lubeck_universal_2004}.
Because the exponents $\mu$ and $\nu$ are linked, the power laws in the
middle row of figure~\ref{fig:LamG} are not independent of the corresponding
power laws in the top row. However, they
rely on different data and hence are not limited by the same finite-size and
finite-distance-from-critical effects.
The power law for the $x$ direction is now seen to be in clear agreement with
the directed percolation exponent, as are those in $t$ and $z$.

Combining these power laws, the laminar gap distributions can be collapsed
  using the relationship $N \ell^\mu = G( \ell \epsilon^\nu)$, where $G$ is an
  unknown function \citep[pp. 111-112]{henkel2008non}.  
In the bottom row of figure~\ref{fig:LamG} we plot our data in collapsing
coordinates using the (2+1)-D percolation exponents.
This collapse is well illustrated by the temporal gap distribution, a
culmination of the excellent fits of $\mu_\parallel$ and
$\nu_\parallel$.  For the $x$-gaps, only gaps of length 100 or greater
are counted, corresponding to the start of the power law in
figure~\ref{fig:LamG}(b).  The collapse of the $z$-gaps is hindered by
the $\nu_\perp$ scaling seen in figure~\ref{fig:LamG}(f), but close to
criticality (last three lines) the data begin to show collapse. 

We have also run simulations in a quasi-1D, streamwise-oriented domain
similar in spirit to the experiments of \cite{lemoult2016directed} (see
the experimental domain in figure \ref{fig:Domain}).  In a domain of size
$[1280h,1.25h,40h]$, the distribution of streamwise laminar gaps near
criticality exhibits a clear power law, in contrast to the poor power-law
behaviour found for streamwise laminar gaps in the full planar system
(figure~\ref{fig:LamG}b). The exponent is $\mu_\perp \simeq 1.748$, as
predicted for systems with a single spatial dimension ((1+1)D directed
  percolation).

  We return to the time evolution shown in figure \ref{fig:Decay} (a).
  Between the evolution at $\Rey<\Rec$ and $\Rey>\Rec$ is the power
  law decay predicted for a directed-percolation process at
  criticality: $F_t \sim t^{-\alpha}$, where
  $\alpha = 2 - \mu_{\parallel} \simeq 0.4505$. Close to criticality,
  we observe evidence for this power law in the data.
This plot highlights a major challenge to simulations near criticality -- well
over $10^5$ time units are required to reach even the moderately small
turbulence fractions simulated here.
As was noted by \citet[p.~32]{Avila_Shear_2013}, these long timescales proved an
  issue in the work of \cite{duguet2010formation}, who in $10^4$ time units of
  simulation were unable to converge turbulence fractions much below 0.4.
  As in the present study, \citet[p.~90]{Avila_Shear_2013} let the system
  evolve for $O(10^6)$ advective time units close to transition.
  Hence both simulation time and domain size can be limiting factors in
  observing the hallmarks of percolation.  
Using directed percolation scalings \citep{takeuchi2009experimental}, the data
above and below criticality collapse onto two curves (figure
\ref{fig:Decay}b), highlighting the universality of directed percolation in
the transition to turbulence.

\section{Discussion}

Over the years
several attempts have been made to quantify the transition to
turbulence and to determine whether or not subcritical shear flows follow the
spatiotemporal scenario of directed percolation.  Such attempts have
consistently been frustrated by the large system sizes required to address the
issue.
Here we have performed simulations of a planar flow of sufficient size that we
have been able to eliminate significant finite-size, finite-time effects,
and thereby to examine in full detail the onset of turbulence in a planar
example.
We have demonstrated both that the equilibrium turbulence fraction increases
continuously from zero above a critical Reynolds number and that statistics of
the turbulent structures exhibit the power-law scalings of the directed
percolation universality class.
%
%
Meeting such demands has necessitated not only turning to the stress-free
boundaries of Waleffe flow, but further truncating the simulations to just
four wall-normal modes.
Performing a comparable computational study directly on plane Couette flow is
currently far beyond available resources.

In light of what we now understand about the scales needed to capture sparse
turbulent structures near the onset of turbulence, we have re-examined the
apparent discontinuous transition to turbulence reported \MC{in} past studies
of plane Couette flow.
The conclusions of those studies were reasonable at that time, but our
  results indicate that prior experimental system sizes were too small, and
  prior simulation times were too short, to accurately capture sustained
  turbulence close to onset -- both space and time constraints can limit
  estimates of true equilibrium dynamics.
Apart from overall issues of scale, we have observed that the scaling
relations in the streamwise and spanwise directions may converge at different
rates. Because shear flows are non-isotropic, it is important to monitor these
directions separately.
These considerations should guide the design of experiments and computations. 
In this regard, it should be noted that while the Reynolds numbers in our
  study differ from those of plane Couette flow, the length and time scales
  are closely comparable to those of plane Couette flow.
The efficiency with which our system can be simulated offers potential for use
in conjunction with future investigation.

While we cannot rule out the possibility that other subcritical shear flows
follow some different route to turbulence, we know that truncated Waleffe flow
contains the essential self-sustaining mechanism of wall-bounded turbulence
and that it produces the oblique turbulent bands that characterize
transitional turbulence in plane Couette and plane channel flow
\citep{chantryTB}.
The closeness of these phenomena suggests that all of these flows exhibit the
same route to turbulence.

\section*{Acknowledgments}

We thank Y.~Duguet and P.~Manneville for useful discussions. We are grateful
to A. Lema\^\i tre for advice on the CML model and specifically for
suggesting the use of $u^*$ as a parameter to vary between discontinuous and
continuous transitions.
M.C. was supported by the grant TRANSFLOW, provided by the Agence Nationale de
la Recherche (ANR).
This research was supported in part by the National Science Foundation under
Grant No. NSF PHY11-25915.
This work was performed using high performance computing resources provided by
the Institut du Developpement et des Ressources en Informatique Scientifique
(IDRIS) of the Centre National de la Recherche Scientifique (CNRS),
coordinated by GENCI (Grand \'Equipement National de Calcul Intensif).

\section*{Appendix. Details of the CML}

We provide here details of the CML model and simulations shown
in \S\ref{sec:CML}. For the most part, these follow previous works
\cite[e.g.][]{bottin1998statistical,Rolf_Directed_1998}.
The local dynamics is given by the map $f$:
\begin{equation}
f(u) = \begin{cases} 
r u, & u \le 1/2 \\
r(1-u) & 1/2 < u \le 1 \\
k (u-u^*) + u^* & 1 < u
\end{cases}
\end{equation}
where $r$, $k$, and $u^*$ are parameters. Here we fix $r=3.0$ and $k=0.8$.
The only difference between this map and those used previously is that here
$u^*$ is a free parameter, rather than being set by the value of $r$ to
$u^*=(r+2)/4$.

The spatial coupling is given by
%
%
%
%
\begin{align*} 
\triangle_f u_{ij} = & \frac{1-\delta}{4}
\left ( f(u_{i-1,j}) - 2f(u_{i,j}) + f(u_{i+1,j}) \right) \\
+ & \frac{1+\delta}{4}
\left ( f(u_{i,j-1}) - 2f(u_{i,j}) + f(u_{i,j+1}) \right)
\end{align*} 
subject to periodic boundary conditions. This term differs from 
the standard coupling
only in that the parameter $\delta$ permits different coupling
strengths in the $i$ and $j$ directions, to mimic the anisotropy of planar
shear flows. We use $\delta = 0.6$.
This anisotropy has no significance for the results presented in this paper
since continuous and discontinuous transitions occur also in the isotropic
case $\delta = 0$.
We show results at $\delta = 0.6$ only for consistency with future
publications.
%

\bibliographystyle{jfm}
\bibliography{biblio}

\begin{thebibliography}{31}
\expandafter\ifx\csname natexlab\endcsname\relax\def\natexlab#1{#1}\fi
\def\au#1{#1} \def\ed#1{#1} \def\yr#1{#1}\def\at#1{#1}\def\jt#1{\textit{#1}}
  \def\bt#1{#1}\def\bvol#1{\textbf{#1}} \def\vol#1{#1} \def\pg#1{#1}
  \def\publ#1{#1}\def\arxiv#1{#1}\def\org#1{#1}\def\st#1{\textit{#1}}

\bibitem[Avila(2013)]{Avila_Shear_2013}
{\sc \au{Avila, K.}} \yr{2013} Shear flow experiments: Characterizing the onset
  of turbulence as a phase transition, {PhD} thesis, {Georg-August} {University
  School of Science}.

\bibitem[Avila {\em et~al.\/}(2011)Avila, Moxey, de~Lozar, Avila, Barkley \&
  Hof]{avila2011onset}
{\sc \au{Avila, K.}, \au{Moxey, D.}, \au{de~Lozar, A.}, \au{Avila, M.},
  \au{Barkley, D.} \& \au{Hof, B.}} \yr{2011}  \at{The onset of turbulence in
  pipe flow}.  \jt{Science}  \bvol{333},  \pg{192--196}.

\bibitem[Barkley(2011)]{barkley11}
{\sc \au{Barkley, D.}} \yr{2011}  \at{Simplifying the complexity of pipe flow}.
   \jt{Phys. Rev. E}  \bvol{84},  \pg{016309}.

\bibitem[Barkley(2016)]{Barkley_Theoretical_2016}
{\sc \au{Barkley, D.}} \yr{2016}  \at{Theoretical perspective on the route to
  turbulence in a pipe}.  \jt{J. Fluid Mech.}  \bvol{803},  \pg{P1}.

\bibitem[Berg{\'e} {\em et~al.\/}(1998)Berg{\'e}, Pomeau \&
  Vidal]{berge1998espace}
{\sc \au{Berg{\'e}, P.}, \au{Pomeau, Y.} \& \au{Vidal, C.}} \yr{1998} {\em
  L'espace {C}haotique\/}.  \publ{Hermann {\'E}d. des {S}ciences et des
  {A}rts}.

\bibitem[Bottin \& Chat{\'e}(1998)]{bottin1998statistical}
{\sc \au{Bottin, S.} \& \au{Chat{\'e}, H.}} \yr{1998}  \at{Statistical analysis
  of the transition to turbulence in plane \protect{Couette} flow}.  \jt{Eur.
  Phys. J. B}  \bvol{6},  \pg{143--155}.

\bibitem[Bottin {\em et~al.\/}(1998)Bottin, Daviaud, Manneville \&
  Dauchot]{bottin1998discontinuous}
{\sc \au{Bottin, S.}, \au{Daviaud, F.}, \au{Manneville, P.} \& \au{Dauchot,
  O.}} \yr{1998}  \at{Discontinuous transition to spatiotemporal intermittency
  in plane \protect{Couette} flow}.  \jt{Europhys. Lett.}  \bvol{43},
  \pg{171--176}.

\bibitem[Chantry {\em et~al.\/}(2016)Chantry, Tuckerman \& Barkley]{chantryTB}
{\sc \au{Chantry, M.}, \au{Tuckerman, L.~S.} \& \au{Barkley, D.}} \yr{2016}
  \at{Turbulent--laminar patterns in shear flows without walls}.  \jt{J. Fluid
  Mech.}  \bvol{791},  \pg{R8}.

\bibitem[Chat{\'e} \& Manneville(1988)]{Chate_Spatio_1988}
{\sc \au{Chat{\'e}, H.} \& \au{Manneville, P.}} \yr{1988}  \at{Spatio-temporal
  intermittency in coupled map lattices}.  \jt{Physica D}  \bvol{32},
  \pg{409--422}.

\bibitem[Duguet {\em et~al.\/}(2010)Duguet, Schlatter \&
  Henningson]{duguet2010formation}
{\sc \au{Duguet, Y.}, \au{Schlatter, P.} \& \au{Henningson, D.~S.}} \yr{2010}
  \at{Formation of turbulent patterns near the onset of transition in plane
  \protect{Couette} flow}.  \jt{J.~Fluid Mech.}  \bvol{650},  \pg{119--129}.

\bibitem[Grassberger(1982)]{grassberger1982phase}
{\sc \au{Grassberger, P.}} \yr{1982}  \at{On phase transitions in
  {S}chl{\"o}gl's second model}.  \jt{Z. Physik B - Condensed Matter}
  \bvol{47},  \pg{365--374}.

\bibitem[Henkel {\em et~al.\/}(2008)Henkel, Hinrichsen, L{\"u}beck \&
  Pleimling]{henkel2008non}
{\sc \au{Henkel, M.}, \au{Hinrichsen, H.}, \au{L{\"u}beck, S.} \&
  \au{Pleimling, M.}} \yr{2008} {\em Non-equilibrium {P}hase {T}ransitions\/},
  ,  \vol{vol.~1}.  \publ{Springer}.

\bibitem[Janssen(1981)]{janssen1981nonequilibrium}
{\sc \au{Janssen, H.-K.}} \yr{1981}  \at{On the nonequilibrium phase transition
  in reaction-diffusion systems with an absorbing stationary state}.  \jt{Z.
  Physik B - Condensed Matter}  \bvol{42},  \pg{151--154}.

\bibitem[Kanazawa {\em et~al.\/}(2017)Kanazawa, Shimizu \&
  Kawahara]{kawahara_private}
{\sc \au{Kanazawa, T.}, \au{Shimizu, M.} \& \au{Kawahara, G.}} \yr{2017}
  Presented at KITP Conference: Recurrence, Self-Organization, and the Dynamics
  of Turbulence, 9-13 January 2017, Kavli Institute for Theoretical Physics.
  http://online.kitp.ucsb.edu/online/transturb-c17/kawahara/.

\bibitem[Kaneko(1985)]{Kaneko_Spatiotemporal_1985}
{\sc \au{Kaneko, K.}} \yr{1985}  \at{Spatiotemporal intermittency in coupled
  map lattices}.  \jt{Progr. Theor. Exp. Phys.}  \bvol{74},  \pg{1033--1044}.

\bibitem[Lemoult {\em et~al.\/}(2016)Lemoult, Shi, Avila, Jalikop, Avila \&
  Hof]{lemoult2016directed}
{\sc \au{Lemoult, G.}, \au{Shi, L.}, \au{Avila, K.}, \au{Jalikop, S.~V.},
  \au{Avila, M.} \& \au{Hof, B.}} \yr{2016}  \at{Directed percolation phase
  transition to sustained turbulence in {C}ouette flow}.  \jt{Nat. Phys.}
  \bvol{12},  \pg{254--258}.

\bibitem[L{\"u}beck(2004)]{Lubeck_universal_2004}
{\sc \au{L{\"u}beck, S.}} \yr{2004}  \at{Universal scaling behavior of
  non-equilibrium phase transitions}.  \jt{Int. J. Mod. Phys. B}  \bvol{18},
  \pg{3977--4118}.

\bibitem[Manneville(2016)]{Manneville_Transition_2016}
{\sc \au{Manneville, P.}} \yr{2016}  \at{Transition to turbulence in
  wall-bounded flows: Where do we stand?}  \jt{Bull. JSME}  \bvol{3},
  \pg{15--00684}.

\bibitem[Marcus \& Lee(1998)]{marcus1998model}
{\sc \au{Marcus, P.} \& \au{Lee, C.}} \yr{1998}  \at{A model for eastward and
  westward jets in laboratory experiments and planetary atmospheres}.
  \jt{Phys. Fluids}  \bvol{10},  \pg{1474--1489}.

\bibitem[Paranjape {\em et~al.\/}(2017)Paranjape, Vasudevan, Duguet \&
  Hof]{duguet_private}
{\sc \au{Paranjape, C.}, \au{Vasudevan, M.}, \au{Duguet, Y.} \& \au{Hof, B.}}
  \yr{2017} Presented at KITP Conference: Recurrence, Self-Organization, and
  the Dynamics of Turbulence, 9-13 January 2017, Kavli Institute for
  Theoretical Physics.
  http://online.kitp.ucsb.edu/online/transturb-c17/hof/rm/jwvideo.html.

\bibitem[Pedlosky(2012)]{pedlosky2012geophysical}
{\sc \au{Pedlosky, J.}} \yr{2012} {\em Geophysical Fluid Dynamics\/}.
  \publ{Springer Science \& Business Media}.

\bibitem[Pomeau(1986)]{pomeau1986front}
{\sc \au{Pomeau, Y.}} \yr{1986}  \at{Front motion, metastability and
  subcritical bifurcations in hydrodynamics}.  \jt{Physica D}  \bvol{23},
  \pg{3--11}.

\bibitem[Prigent {\em et~al.\/}(2003)Prigent, Gr{\'e}goire, Chat{\'e} \&
  Dauchot]{prigent2003long}
{\sc \au{Prigent, A.}, \au{Gr{\'e}goire, G.}, \au{Chat{\'e}, H.} \&
  \au{Dauchot, O.}} \yr{2003}  \at{Long-wavelength modulation of turbulent
  shear flows}.  \jt{Physica D}  \bvol{174},  \pg{100--113}.

\bibitem[Rolf {\em et~al.\/}(1998)Rolf, Bohr \& Jensen]{Rolf_Directed_1998}
{\sc \au{Rolf, J.}, \au{Bohr, T.} \& \au{Jensen, M.}} \yr{1998}  \at{Directed
  percolation universality in asynchronous evolution of spatiotemporal
  intermittency}.  \jt{Phys. Rev. E}  \bvol{57},  \pg{R2503}.

\bibitem[Sano \& Tamai(2016)]{sano2016universal}
{\sc \au{Sano, M.} \& \au{Tamai, K.}} \yr{2016}  \at{A universal transition to
  turbulence in channel flow}.  \jt{Nat. Phys.}  \bvol{12},  \pg{249--253}.

\bibitem[Shih {\em et~al.\/}(2016)Shih, Hsieh \&
  Goldenfeld]{Shih_Ecological_2016}
{\sc \au{Shih, H.}, \au{Hsieh, T.} \& \au{Goldenfeld, N.}} \yr{2016}
  \at{Ecological collapse and the emergence of travelling waves at the onset of
  shear turbulence}.  \jt{Nat. Phys.}  \bvol{12},  \pg{245--248}.

\bibitem[Suri {\em et~al.\/}(2014)Suri, Tithof, Mitchell~Jr, Grigoriev \&
  Schatz]{suri2014velocity}
{\sc \au{Suri, B.}, \au{Tithof, J.}, \au{Mitchell~Jr, R.}, \au{Grigoriev,
  R.~O.} \& \au{Schatz, M.~F.}} \yr{2014}  \at{Velocity profile in a two-layer
  {K}olmogorov-like flow}.  \jt{Phys. Fluids}  \bvol{26},  \pg{053601}.

\bibitem[Takeuchi {\em et~al.\/}(2007)Takeuchi, Kuroda, Chat{\'e} \&
  Sano]{takeuchi2007directed}
{\sc \au{Takeuchi, K.~A.}, \au{Kuroda, M.}, \au{Chat{\'e}, H.} \& \au{Sano,
  M.}} \yr{2007}  \at{Directed percolation criticality in turbulent liquid
  crystals}.  \jt{Phys. Rev. Lett.}  \bvol{99},  \pg{234503}.

\bibitem[Takeuchi {\em et~al.\/}(2009)Takeuchi, Kuroda, Chat{\'e} \&
  Sano]{takeuchi2009experimental}
{\sc \au{Takeuchi, K.~A.}, \au{Kuroda, M.}, \au{Chat{\'e}, H.} \& \au{Sano,
  M.}} \yr{2009}  \at{Experimental realization of directed percolation
  criticality in turbulent liquid crystals}.  \jt{Phys. Rev. E}  \bvol{80},
  \pg{051116}.

\bibitem[Tsukahara \& Ishida(2017)]{tsukahara_private}
{\sc \au{Tsukahara, T.} \& \au{Ishida, T.}} \yr{2017} private communication.

\bibitem[Xiong {\em et~al.\/}(2015)Xiong, Tao, Chen \&
  Brandt]{xiong2015turbulent}
{\sc \au{Xiong, X.}, \au{Tao, J.}, \au{Chen, S.} \& \au{Brandt, L.}} \yr{2015}
  \at{\protect{Turbulent bands in plane-Poiseuille flow at moderate Reynolds
  numbers}}.  \jt{Phys. Fluids}  \bvol{27},  \pg{041702}.

\end{thebibliography}

\end{document}